%% file: main.tex
\pdfoutput=1
\documentclass{article}

\usepackage[letterpaper,top=2cm,bottom=2cm,left=3cm,right=3cm,marginparwidth=1.75cm]{geometry}
\usepackage{titletoc}

\usepackage{authblk}
\usepackage{amsmath}
\usepackage{graphicx}
\usepackage[numbers, sort, compress]{natbib}
\usepackage[dvipsnames]{xcolor}
\usepackage[utf8]{inputenc} 
\usepackage[T1]{fontenc}    
\usepackage[hyphens]{url}
\usepackage{hyperref}     
\PassOptionsToPackage{url}{hyphens} 
\hypersetup{colorlinks=true,allcolors=darkblue}
\usepackage[hyphenbreaks]{breakurl}

\usepackage{booktabs}       
\usepackage{amsfonts}       
\usepackage{nicefrac}       
\usepackage{listings}
\usepackage{microtype}      
\usepackage{mathtools}
\usepackage{refcount}
\usepackage{algorithm}
\usepackage{algpseudocode}
\usepackage{amsmath,amssymb, amsthm}
\usepackage{physics}
\usepackage[subtle, mathdisplays=tight, charwidths=normal, leading=normal]{savetrees}
\usepackage{caption}
\usepackage[bottom]{footmisc}
\usepackage{subcaption}
\usepackage{float}
\usepackage{enumitem}
\usepackage{tocloft}
\usepackage{letltxmacro}
\usepackage{xparse}
\usepackage{xcolor}
\usepackage{savetrees}

\definecolor{glosscolor}{HTML}{427842}
\definecolor{glosslinkcolor}{HTML}{56ae57}

\definecolor{darkblue}{HTML}{0d75f8}

\newcommand{\notocsection}[1]{%
\refstepcounter{section}%
\section*{\thesection \quad #1}}%

\newcommand{\notocsubsection}[1]{%
\refstepcounter{subsection}%
\subsection*{\thesubsection \quad #1}}%

\providecommand{\tightlist}{%
  \setlength{\itemsep}{0pt}\setlength{\parskip}{0pt}}
  
\newcommand{\definedterm}[1]{\textbf{#1}}

\renewcommand\Affilfont{\fontsize{10}{10.8}\itshape}

\makeatletter
\renewcommand\AB@affilsepx{, \protect\Affilfont}
\makeatother

\newcommand{\glosslink}[2]{\hyperlink{#1}{\textbf{\textcolor{glosslinkcolor}{#2}}}}

\newcommand*\samethanks[1][\value{footnote}]{\footnotemark[#1]}

\newcommand{\custompar}[1]{\vspace{.2cm}\noindent{\bf #1.}\:}

\title{\textrm{Report of the 1st Workshop on Generative AI and Law}}
\author[1, 2, 3]{A. Feder Cooper\thanks{Equal contribution. Correspondence: \texttt{genlaw.org@gmail.com}. We thank Matthew Jagielski, Andreas Terzis, and Jonathan Zittrain for feedback on this report. We thank Google, Microsoft, Schmidt Futures, OpenAI, Anthropic, Cornell Law School, and ML Collective for their generous sponsorship.}}
\author[1, 2, 4]{Katherine Lee\samethanks}
\author[1, 5, 6]{James Grimmelmann\samethanks}\looseness=-1
\author[4, 7]{Daphne Ippolito\samethanks}

\author[8]{\\Christopher Callison-Burch}
\author[4]{Christopher A. Choquette-Choo}
\author[1, 9]{Niloofar Mireshghallah}
\author[10]{Miles Brundage}
\author[1, 2]{David Mimno}
\author[1, 2, 5]{Madiha Zahrah Choksi}

\author[11]{Jack M. Balkin}
\author[4]{Nicholas Carlini}
\author[2]{Christopher De Sa}
\author[12]{Jonathan Frankle}
\author[1, 13]{Deep Ganguli}
\author[14]{Bryant Gipson}
\author[15]{Andres Guadamuz}
\author[16]{Swee Leng Harris}
\author[17]{Abigail Z. Jacobs}
\author[18]{Elizabeth Joh}
\author[19]{Gautam Kamath}
\author[20]{Mark Lemley}
\author[21]{Cass Matthews}
\author[10]{Christine McLeavey}
\author[22]{Corynne McSherry}
\author[3]{Milad Nasr}
\author[23]{Paul Ohm}
\author[4]{Adam Roberts}
\author[10]{Tom Rubin}
\author[24]{Pamela Samuelson}
\author[1]{Ludwig Schubert}
\author[25]{Kristen Vaccaro}
\author[26]{Luis Villa}
\author[27]{Felix Wu}
\author[28]{Elana Zeide}

\affil[1]{GenLaw Organizers}
\affil[2]{Cornell University}
\affil[3]{Google Research}
\affil[4]{Google DeepMind}
\affil[5]{Cornell Tech}
\affil[6]{Cornell Law School}
\affil[7]{Carnegie Mellon University}
\affil[8]{University of Pennsylvania}
\affil[9]{University of Washington}
\affil[10]{OpenAI}
\affil[11]{Yale Law School}
\affil[12]{DataBricks (MosaicML)}
\affil[13]{Anthropic}
\affil[14]{Google}
\affil[15]{University of Sussex}
\affil[16]{Luminate Group}
\affil[17]{University of Michigan}
\affil[18]{U.C. Davis, School of Law}
\affil[19]{University of Waterloo}
\affil[20]{Stanford Law School}
\affil[21]{Microsoft}
\affil[22]{Electronic Frontier Foundation}
\affil[23]{Georgetown Law School}
\affil[24]{U.C. Berkeley, School of Law}
\affil[25]{U.C. San Diego}
\affil[26]{Tidelift}
\affil[27]{Cardozo Law}
\affil[28]{Nebraska College of Law}

\renewcommand*{\Affilfont}{\normalsize}

\date{}

\makeatletter
\renewcommand\AB@affilsepx{, \protect\Affilfont}
\makeatother

\newcounter{UntUntKap}[section]  
\renewcommand\theUntUntKap{\thesection.\arabic{UntUntKap}}%

\makeatletter
\AtBeginDocument{\let\l@UntUntKap=\l@subsubsection}%
\newcommand\CheckWhetherArgumentIsEmpty[1]{%
  \romannumeral0%
  \ifcat A\detokenize{#1}A%
    \@firstoftwo{\expandafter\expandafter\expandafter}{} %
    \expandafter\@firstoftwo
  \else
    \@firstoftwo{\expandafter\expandafter\expandafter}{} %
    \expandafter\@secondoftwo
  \fi
}%
\newcommand\nummeriere{%
  \@ifstar\nummeriere@star\nummeriere@nostar
}%
\newcommand{\nummeriere@star}[1][]{%
    \refstepcounter{UntUntKap}%
    \addcontentsline{toc}{UntUntKap}{%
      \CheckWhetherArgumentIsEmpty{#1}{\protect\kern-.6em}{#1}%
    }
  \@bsphack\@esphack
}%
\newcommand{\nummeriere@nostar}[1][]{
    \refstepcounter{UntUntKap}%
    \addcontentsline{toc}{UntUntKap}{%
      \ifnum2>\c@secnumdepth\else
        \protect\numberline{\theUntUntKap}%
      \fi
      \CheckWhetherArgumentIsEmpty{#1}{\protect\kern-.6em}{#1}%
    }
  \@bsphack\@esphack
}

\makeatother

\begin{document}
\maketitle

\input{section/00-abstract}


\newpage
\input{section/10-introduction}

\input{section/15-vision}
\input{section/20-shared}
\input{section/30-unique}

\input{section/40-taxonomy}
\input{section/50-agenda}
\input{section/60-conclusion}

\newpage
\bibliographystyle{plainnat}
\bibliography{references}

\newpage 
\appendix

\input{section/99-appendix}

\end{document}

%% file: section/00-abstract.tex
\begin{abstract}
This report presents the takeaways of the inaugural Workshop on Generative AI and Law (GenLaw), held in July 2023. A cross-disciplinary group of practitioners and scholars from computer science and law convened to discuss the technical, doctrinal, and policy challenges presented by law for Generative AI, and by Generative AI for law, with an emphasis on U.S. law in particular. 
We begin the report with a high-level statement about why Generative AI is both immensely significant and immensely challenging for law. 
To meet these challenges, we conclude that there is an essential need for 1) a \definedterm{shared knowledge base} that provides a common conceptual language for experts across disciplines; 2) clarification of the distinctive \definedterm{technical capabilities} of generative-AI systems, as compared and contrasted to other computer and AI systems; 3) a logical taxonomy of the \definedterm{legal issues} these systems raise; and, 4) a concrete \definedterm{research agenda} to promote collaboration and knowledge-sharing on emerging issues at the intersection of Generative AI and law. 
In this report, we synthesize the key takeaways from the GenLaw workshop that begin to address these needs. 
All of the listed authors contributed to the workshop upon which this report is based, but they and their organizations do not necessarily endorse all of the specific claims in this report. 
\end{abstract}

\vspace{1in}
\noindent This work is licensed under CC BY 4.0. To view a copy of this license, visit \url{http://creativecommons.org/licenses/by/4.0/}. 

%% file: section/10-introduction.tex
\notocsection{Introduction}\label{sec:intro}
\vspace{-.1cm}

The inaugural Generative AI and Law (GenLaw) workshop took place on July 29th and 30th, in Honolulu, Hawai'i, where it was co-located with the 40th International Conference on Machine Learning (ICML).
The workshop was organized in response to the intense interest in (and scrutiny of) recent public advancements in generative-AI technology. 
The primary goal was to bring together experts in machine learning (ML) and law to discuss the legal challenges that generative-AI technology raises. 
To promote concrete and focused discussions, we chose to make intellectual property (IP) and privacy the principal legal topics of the first workshop.
Other significant topics discussed included free speech, products liability, and transparency.
For this first workshop, most discussion was limited to considerations of U.S. law. 

The workshop was convened over two days. 
The first day (July 29th) was a public session held as part of ICML, consisting of keynote lectures, panel discussions, lightning talks, and a poster session, all dealing with research issues at the intersection of Generative AI and law. 
The second day (July 30th), was 
held off-site, at which approximately forty participants conducted a series of roundtable discussions to dig deeper into  significant issues identified on the first day.

This report reflects the takeaways from the roundtable discussions.
They are organized into five broad headings, reflecting the participants' consensus about the most urgently needed contributions to the research area of Generative AI and law:
\begin{enumerate}[itemsep=.1cm]
    \item A high-level statement about why Generative AI is both immensely significant and immensely challenging for law (Section~\ref{sec:vision});
    \item The beginnings of a shared knowledge base that provides a common conceptual language for experts across disciplines (Section~\ref{sec:knowledge}); 
    \item Clarification of the unique capabilities and issues of generative-AI systems, setting them in relation to the broader landscape of artificial-intelligence and machine-learning technologies (Section~\ref{sec:challenges});
    \item An initial taxonomy of the legal issues at play (Section~\ref{sec:taxonomy}); and,
    \item A concrete research agenda to promote collaboration and progress on emerging issues at the intersection of Generative AI and law (Section~\ref{sec:agenda}). 
\end{enumerate}

To best serve these ends, this report does not delve into the technical details of specific generative-AI systems, the legal details of complaints and lawsuits involving those systems, or policy proposals for regulators. 
Our intended audience is scholars and practitioners who are already interested in engaging with issues at the intersection of Generative AI and law, for example, ML researchers who have familiarity with some of the ongoing lawsuits regarding Generative AI, and lawyers who have familiarity with terms like ``large language model.'' 
We focus our attention on synthesizing reflections from the workshop to highlight key issues that need to be addressed for successful research progress in this emerging and fundamentally interdisciplinary area. 



%% file: section/15-vision.tex
\vspace{-.25cm}
\notocsection{The Impact of Generative AI on Law}\label{sec:vision}
\vspace{-.1cm}

Generative AI is ``generative'' because it generates text, images, audio, or other types of output. 
But it is also ``generative'' in the sense of Jonathan Zittrain's theory of generative technologies: it has the ``capacity to produce unanticipated change through unfiltered contributions from broad and varied audiences''~\cite{zittrainfuture}. 
As a result, generative-AI systems will be both immensely societally significant --- too significant for governments to ignore or to delay dealing with --- and present an immensely broad range of legal issues.
To see why, it is useful to consider \citet{zittrainfuture}'s five dimensions of generativity: 
\begin{description}[itemsep=.1cm, topsep=.1cm]
	\item[Leverage] \textit{A technology provides leverage when it makes difficult tasks easier.} 
	Generative AI is widely recognized for its use in creativity; programming; retrieving and synthesizing complex bodies of knowledge; and automating repetitive tasks.
	\item[Adaptability] \textit{A technology is adaptable when it can be applied to a wide range of uses.} 
	Generative AI is celebrated for its adaptability. It has been applied to programming, painting, language translation, drug discovery, fiction, educational testing, graphic design, and much more.
	\item[Ease of mastery] \textit{A technology is easy to master when users without specialized training can readily adopt and adapt it.} 
	While some generative-AI methodologies, such as model pre-training, still require technical skills, the ability to use chat-style, interactive, natural-language prompting to control generative-AI systems greatly reduces the difficulty of adoption. 
	Users without programming or ML backgrounds have been able to use Generative AI for numerous tasks.
	\item[Accessibility] \textit{A technology is accessible when there are few barriers to its use.} 
	Cost is the most obvious barrier, but other barriers can include regulation, secrecy, and linguistic limits. 
    The creation of cutting-edge generative-AI models from scratch requires  enormous inputs of data, compute, and human expertise --- currently limiting model creation to a handful of institutions --- but services allowing inference with these models are widely available to the public.
    These services are inexpensive for users making small numbers of queries, and they tend to operate close to real-time, making generative outputs (at least appear) low-cost to produce, in terms of time. 
	\item[Transferability] \textit{A technology is transferable when changes in it can easily be conveyed to others.} 
	Once pre-trained or fine-tuned, generative-AI models can be easily shared, prompts and prompting techniques are trivially easy to describe, and systems built around generative-AI models can be made broadly available at increasingly low effort and cost.
\end{description}
In short, Generative AI hits the generativity jackpot. 
It provides enormous leverage across a wide range of tasks, is readily built on by a huge range of users, and facilitates rapid iterative improvement as those users share their innovations with each other.

\citet{zittrainfuture}'s two examples of supremely generative technologies from 2008 are computers and the Internet. 
Generative AI seems likely to be a third. 
No other technology of the last two decades even comes close. 
Regulators and legal scholars should expect that Generative AI will raise legal and policy challenges that are comparable in scope, scale, and complexity to those raised by computers and the Internet.

The Internet-law analogy also provides guidance on how technologists and lawyers can approach this shared challenge.
They must have a common vocabulary so that their contributions are mutually intelligible (Section~\ref{sec:knowledge}).
Lawyers must have a sufficient foundation of technical understanding to be able to apply their expertise in law accurately (Section~\ref{sec:challenges}). 
Technologists, for their part, must have a sufficient foundation of legal knowledge to identify legally significant technical interventions (Section~\ref{sec:taxonomy}).
And both groups need a common research agenda to collaborate and iterate rapidly on effective projects that advance a shared understanding how Generative AI and the legal system interact (Section~\ref{sec:agenda}).
The aim of this report is to lay down a starting framework for these tasks.

%% file: section/20-shared.tex
\vspace{-.25cm}
\notocsection{Developing a Shared Knowledge Base}\label{sec:knowledge}
\vspace{-.1cm}

It became apparent over the course of the GenLaw roundtable discussions that some commonly used terms have different meanings in machine learning and in law.
Sometimes, both groups have been working to develop deep understandings of important but hard-to-capture concepts.
The term \textit{privacy} is a prominent such example.
Technologists' formal definitions (such as \glosslink{gloss:Differential Privacy}{differential privacy}) do not always encompass the wide range of interests protected by privacy law; similarly, it can be hard to put the holistic definitions used by legal scholars into computationally tractable forms that can be deployed in actual systems.\footnote{The U.S. Census has sparked debate over its use of \glosslink{gloss:Differential Privacy}{differential privacy}: a technique that provides strong theoretical guarantees of privacy preservation. Critics question whether or not the definition of privacy reflected in differential privacy accords with the census's broader goals of privacy preservation. Differential privacy is also sometimes used in Generative AI, though it is context-dependent whether its definition of ``privacy'' is meaningful for generative-AI applications~\cite{brown2022privacy}.}
These communication barriers are real, but, as was clear during GenLaw, the two communities have generally understood that they mean something different by ``privacy'' and have read each others' work with a working understanding of these differences in mind.

We also observed, however, that there are also various types of misunderstandings in terminology across disciplines. 
At GenLaw, some terms were used in different ways because members of one community did not even realize a loosely-defined term in their community was a term of art in the other, or because the meaning they assumed a term had was subtly different from how it was actually used in writing.
These conflicting definitions and translation gaps hampered our ability to collaborate on assessing emerging issues.
For example, technologists use the term \glosslink{gloss:Pre-Training and Fine-Tuning}{pre-training} to refer to an early, general-purpose phase of the model training process, but legal scholars assumed that the term referred to a \glosslink{gloss:Data Curation and Pre-Processing}{data preparation stage} prior to and independent of training. 
Similarly, many technologists were not aware of the importance of \glosslink{gloss:Harm}{harms} as a specific and consequential concept in law, rather than a general, non-specific notion of unfavorable outcomes.
We found our way to common understandings only over the course of our conversations, and often only after many false starts.

Thus, first and foremost, we need to have a shared understanding of baseline concepts in both Generative AI and law. 
Even when it is not possible to pin down terms with complete precision, it is important to have clarity about which terms are ambiguous or overloaded.
We believe that there three significant ways that computer scientists and legal scholars can contribute to creating this shared understanding:
\begin{enumerate}[itemsep=.1cm, topsep=.1cm]
    \item They can build \definedterm{glossaries} of definitions of important terms in machine learning and law, which can serve both as textbooks and as references  (Section~\ref{sec:glossary}). Throughout this piece, glossary terms are identified in \textcolor{glosslinkcolor}{\textbf{green}} and serve as links to the corresponding glossary entry in the Appendix.
    \item They can develop well-crafted \definedterm{metaphors} to clarify complex concepts across disciplinary boundaries.  Even imperfect metaphors are useful, as they can serve to highlight where concepts in Generative AI deviate from intuitions that draw on more traditional examples  (Section~\ref{sec:metaphors}). 
    \item They can \definedterm{keep current} with the state of the art, and help others to do so. This does not just mean the fundamentals of machine learning and Generative AI (although these are certainly important). It also means being alert to the plethora of ways that generative-AI systems are be deployed in practice, and the commonalities and differences between these systems (Section~\ref{sec:business}).
\end{enumerate}

\vspace{-.2cm}
\notocsubsection{Identifying and Defining Terms}\label{sec:glossary}

The GenLaw organizers and participants have collaborated to create an initial glossary of important terms at the intersection of law and Generative AI. (It follows as Appendix~\ref{app:glossary}.)
The glossary has two primary goals.

First, it identifies terms of art with technical or multiple meanings.
Both law and machine learning commonly give specific, technical definitions to words that also have general, colloquial meanings, such as ``attention'' or ``harm.''\footnote{See the entry for \glosslink{gloss:Attention}{attention} in the glossary for the machine-learning definition.}
Sometimes, the redefinition runs the other way, when a technical term has taken on a broader meaning in society at large.
To technologists, an \glosslink{gloss:Algorithm}{algorithm} is simply a precise rule for carrying out a procedure, but the term has come to be popularly associated with specific technologies for ranking social media posts based on expected interest.
The words ``goal'' and ``objective'' are sometimes used interchangeably in English, but \glosslink{gloss:Objective}{objective} is a term of art in machine learning, describing a mathematical function used during training. 
On the other hand, ``goal'' does not have an agree-upon technical definition. In machine learning, it is typically used colloquially to describe our overarching desire for model behaviors, which cannot be written directly in math. 

Second, the glossary provides succinct definitions of the most critical concepts for a non-expert in either law or ML (or both).
These definitions are \textit{not} intended to cover the full complexity of a concept or term from the expert perspective.
For example, one could write volumes on privacy --- and many have, arguably for thousands of years.\footnote{This is why we do not attempt a definition of privacy in Appendix~\ref{app:glossary}.}  
Instead, our purpose here is simply to show technologists that there is more to privacy than removing \glosslink{gloss:Personally Identifiable Information (PII)}{personally identifiable information (PII)}.

The glossary is offered as a starting point, not a finish line.
The field is in flux; its terminology will evolve as new technologies and controversies emerge.
We will host and update this glossary on the GenLaw website.\footnote{See \url{https://genlaw.github.io/glossary.html}.}
We hope that these definitions will serve as a baseline for more effective communication across disciplines about emerging issues.

\vspace{-.2cm}
\notocsubsection{Crafting Useful Metaphors}\label{sec:metaphors}


Well-chosen metaphors can provide a useful mental model for thinking through complex concepts. Metaphors are also widely used in both machine learning and law. 
For machine-learning practitioners, these metaphors can also be sources of inspiration; the  idea of an ``artificial neural network'' was inspired by the biology of neurons in the human brain \citep{boers1993biological}.
Analogy and metaphor are central to legal rhetoric~\citep{solove2001metaphor}; they provide a  rational framework for thinking through the relevant similarities and differences between cases.
Metaphors, however, can also simplify and distort;  nevertheless, understanding the ways that a metaphor fails to correctly describe a concept can still be instructive in helping to clarify one's thinking.


At the GenLaw workshop, we discussed instructive metaphors for Generative AI extensively.
We give two examples from this discussion here (anthropomorphism and memorization) and describe additional ones in Appendix~\ref{app:metaphors}.



\custompar{Metaphorical anthropomorphism}
Metaphorical anthropomorphism is the personification of a non-human entity; it applies metaphors that compare the entity's traits to human characteristics, emotions, and behaviors. 
Machine-learning practitioners commonly use terms that  anthropomorphize machine-learning models, for example, saying models ``learn,'' ``respond,'' ``memorize,'' or ``hallucinate.''
Such metaphors can lead people to conclude that a machine-learning system is completing such actions using the same mechanisms and thought processes that a human would.
However, while practitioners may sometimes be inspired in their designs by biological phenomena (e.g., \glosslink{gloss:Neural Network}{neural networks} contain ``neurons'' that ``fire'' analogously to those in the human brain), they by and large do not mean that machine-learning models ``learn'' or ``memorize'' in exactly the same way that humans complete these actions. 
Instead, these should be considered terms of art --- perhaps inspired by human actions, but grounded in technical definitions that bear little resemblance to human mechanisms.

Some within the GenLaw community have advocated for using different terms to describe these processes --- ones that do not elicit such strong comparisons to human behavior~\citep[and citations therein]{cooper2022accountability}.
However, until we have better terms, understanding when a term is indeed a term of art, and the ways that it is inspired by (but not equivalent to) colloquial understandings, will remain a critical part of any interdisciplinary endeavour. 

\custompar{Memorization}
The terms \glosslink{gloss:Memorization}{memorization} and \glosslink{gloss:Regurgitation}{regurgitation} are very common in the machine-learning literature.
Roughly speaking, memorization and regurgitation can be treated interchangeably. They both signify when a machine-learning model encodes details from its training data, such it is capable of generating outputs that closely resemble its training data.\footnote{Some differentiate ``memorization'' and ``regurgitation,'' with regurgitation referring to a model's ability to output its training data (via generation) and memorization referring to a model \textit{containing} a perfect copy of its training data (regardless of whether or not it is regurgitated).
In practice, the two terms are commonly used interchangeably. 
} 
The machine-learning experts who coined these terms created precise definitions that can be translated into quantified metrics; 
these definitions refer to specific ways to measure the amount of memorization (as it is technically definition) present in a model or its outputs~\citep{carlini2023quantifying, ippolito2023preventing,anil2023palm,kudugunta2023madlad}.

Unfortunately, the connection to the colloquial meanings of these words can cause confusion. 
Some outside of the machine-learning community misinterpet ``Generative AI memorization'' to include functionality that goes beyond what machine-learning practitioners are actually measuring with their precise definitions.
One such example is discussions around text-to-image generative models ``memorizing'' an artist's style \citep{newyorker-stealing}. 
While measuring stylistic similarity is an active area of machine-learning research~\citep{casper2023measuring}, it is \emph{not} equivalent to \glosslink{gloss:Memorization}{memorization} in the technical sense of the word. 

Other misunderstandings can arise due to the fact that ``memorization'' and ``regurgitation'' are imperfect analogies for the underlying processes that machine-learning scientists are measuring. 
People deliberately memorize; for example, an actor will actively commit a script to memory.
In contrast, 
models do not deliberately memorize; the training examples that end up memorized by a model were not treated any differently during training than the ones that were not memorized.\footnote{Models are trained using an \glosslink{gloss:Objective}{objective function} that \glosslink{gloss:Loss}{rewards} the model for producing data that looks similar to the training data during the training phase. However, the \textit{goal} of model training is not to reproduce the training data, so other techniques, such as regularization or alignment are used to help a model \glosslink{gloss:Generalization}{generalize}. See Section \ref{sec:glossary} for a discussion on ``goal'' versus \glosslink{gloss:Objective}{objective} in machine learning.
}
Importantly, humans distinguish between memorizing (which has intentionality) and ``remembering'' (when a detail is recalled without the intent to memorize it).
Generative-AI models have no such distinction. 
For humans, it is how we personally feel about a thought, action, or vocalization that leads us to call it ``memorized.'' But for models, which lack intent or feeling, \glosslink{gloss:Memorization}{memorization} is merely a property assigned to their outputs and \glosslink{gloss:Weights}{weights} through technical definitions. 

\vspace{-.1cm}
\notocsubsection{Understanding Evolving Business Models}\label{sec:business} 

Developing glossaries and metaphors can help with staying current about generative-AI technology, but they do not necessarily capture all of the different ways that generative-AI functionality can be put to use in practice. 
To understand real-world uses, it is also important to have a working understanding of different generative-AI production processes. 

There are a variety of evolving business models that have yet to solidify into definitive patterns. In this uncertain environment, myths and folk wisdom can proliferate. 
Real, current information about business models can be very useful for understanding who is involved in the production, maintenance, and use of different parts of generative-AI systems. Business models prove useful for appreciating the quantity and diversity of actors (not just technologies) that enable generative-AI functionality. 

We highlight four patterns from discussion at GenLaw:
\begin{enumerate}[itemsep=.1cm, topsep=.1cm]
    \item \textbf{Business-to-consumer (B2C) hosted services}: Several companies have released direct-to-consumer applications and \glosslink{gloss:Application Programming Interface (API)}{application programming interfaces (APIs)} for producing generations. For example, on the large end of the spectrum, there are chatbots powered by main players in language modeling (e.g., OpenAI's ChatGPT, Anthropic's Claude, Google's Bard, etc.). There are also smaller companies that have released similar tools (e.g., Midjourney's~\citep{midjourney} or Ideogram's~\citep{ideogram} text-to-image generation applications).  These companies provide a mix of entry points to their systems and models, including user interfaces and APIs, often offered via subscription-based services. Typically, the systems and models developed by these companies are hosted in proprietary services; users can access these services to produce generations but, with some notable exceptions (e.g., fine-tuning APIs), cannot directly alter or interact with the models embedded within them.
    \item \textbf{Business-to-business (B2B) integration with hosted services}: Other business models allow for businesses to integrate generative-AI functionality into their products either via direct partnership/integration or through the use of \glosslink{gloss:Application Programming Interface (API)}{APIs}. For example, ChatGPT functionality is integrated into Microsoft Bing search (via a close partnership between Microsoft and OpenAI). Poe is developed based on a partnership between Anthropic and Quora~\citep{poe}.\footnote{We cannot tell from press releases whether Poe also uses the API or if their partnership is of a different nature. Often, we will not be able to tell the nature of the business relationship (unless disclosed publicly) between corporate partners.} 
    In other cases, companies develop generative-AI products by becoming corporate customers of generative-AI-company APIs, as opposed to bespoke business partners. These types of business relationships can either lead to developing new features for existing products, or new products altogether. 
    \item \textbf{Products derived from open models and datasets}: The business models discussed above depend on \glosslink{gloss:Closed Software}{proprietary systems}, \glosslink{gloss:Closed Model}{models}, and/or \glosslink{gloss:Closed Dataset}{datasets}. Other options include (or directly rely on) \glosslink{gloss:Open Software}{open-source software}, \glosslink{gloss:Open Model}{models}, and \glosslink{gloss:Open Dataset}{datasets}, which can be downloaded and put to use (e.g., training new models or fine-tuning existing model \glosslink{gloss:Checkpoint}{checkpoints}). Some companies operate distinctly (or partially) with open-source product offerings, such as some versions of Stable Diffusion~\citep{rombach2022diffusion} offered by Stability AI~\citep{stability}. Others operate in a mixed fashion, for example, individuals can openly download Meta's Llama-model family (the different models' \glosslink{gloss:Weights}{weights}), but the details of the training data are closed~\citep{llama2}. 
    \item \textbf{Companies that operate at specific points in the generative-AI supply chain}: Any link (or subset of links) in the \glosslink{gloss:Supply Chain}{supply chain}~\citep{lee2023talkin} could potentially become a site of specific business engagement with generative-AI technology. For those interested in issues at the intersection of Generative AI and law, it will be important not only to be familiar with the roles and scope of companies in these areas, but also how they interact and inter-operate. We provide three examples below of emerging sites of business engagement. 
    \begin{itemize}[leftmargin=.5cm, topsep=1pt, itemsep=1pt] 
        \item \emph{Datasets}: There may be companies that engage only with the dataset collection and curation aspects of Generative AI (similar to how data brokers function in other industries). Scale AI~\citep{scaleai} is one such company that works on data \glosslink{gloss:Examples}{example} annotation for generative-AI training datasets. 
        \item \emph{Training diagnostics}: There are some companies that handle aspects of data analysis and diagnostics for generative-AI-model training dynamics, like Weights \& Biases~\citep{weights-biases}. 
        \item \emph{Training and deployment}: While it is generally tremendously costly to train and deploy large generative-AI models, advancements in open-source technology and at smaller companies (in both software and hardware) have helped make training custom models more efficient and affordable. There are now several companies that develop solutions for bespoke model training and serving, such as MosaicML (acquired by DataBricks)~\citep{mosaic} and Together AI~\citep{together}. 
    \end{itemize} 
\end{enumerate}

We defer additional discussion of \glosslink{gloss:Open Software}{open}- vs. \glosslink{gloss:Closed Software}{closed}-source software to Appendix~\ref{app:glossary}. The important point that we want to highlight here is that there are many ways that generative-AI models may be integrated into software systems, and that there are many different types of business models associated with the training and use of these models. This landscape is likely to continue to evolve as new business players enter the field. 

%% file: section/30-unique.tex
\vspace{-.25cm}
\notocsection{Pinpointing Unique Aspects of Generative AI}\label{sec:challenges}

During the roundtable discussions, the legal scholars and practitioners had a recurring question for the machine-learning experts in the room: What's so special about Generative AI? 
Clearly, the outputs created by Generative AI today are better than anything we have seen before, but what is the ``magic'' that makes this the case?
It became clear that answering this question, even just in broad strokes, could be useful for providing more precise analysis of the legal issues at play (Section~\ref{sec:taxonomy}). 
In this section, we summarize three aspects of Generative AI for which it can be productive to consider recent developments in AI as meaningfully novel or different in comparison to past technology. 
These include (1) the transition from training models to perform narrowly-defined tasks to training them for open-ended ones (Section~\ref{sec:openended}), (2) the role of the modern \glosslink{gloss:Pre-Training and Fine-Tuning}{training data} (Section~\ref{sec:dividing}) and generative-AI pipelines (Section~\ref{sec:system}) and (3) how the scaling up of pre-existing techniques has enabled the quality and variety we see in generations today (Section~\ref{sec:scale})

\vspace{-.2cm}
\notocsubsection{The Transition to Very Flexible Generative Models}\label{sec:openended}
\vspace{-.1cm}

In the past, machine-learning models tended to be trained to perform narrowly defined discriminative tasks, such as labeling an image according to the class of object it depicts \cite{deng2012mnist,deng2009imagenet} or classifying the sentiment of a sentence as positive or negative \citep{kiritchenko2014sentiment}.
Modern generative-AI models change this paradigm in two ways.\footnote{For a longer summary of this transition from task-specific, discriminative models to generative models in Generative AI, see Parts I.A and I.B of \citet{lee2023talkin}.}

First, there has been a shift from discriminative models, which have simple outputs like a class label (e.g., \texttt{dog} or \texttt{cat} for an image classifier), to generative models, which output complex content, such as entire images or paragraphs of text (e.g., given the input of \texttt{cat}, outputting a novel image of a cat, sampling from the near-infinite space of reasonable cat images it could create)~\citep[Part I.A]{lee2023talkin}.\footnote{As we discuss at the top of Section~\ref{sec:taxonomy}, today's generative-AI models are used to solve both discriminative tasks
(e.g., sentiment classification)
and generative tasks that result in expressive content
(e.g., producing paragraphs of text). 
}

Second, there has been a shift toward using single, general-purpose models to solve many different tasks, rather than employing a model customized to each task we would like to perform.
Even a few years ago, it was common to take a \glosslink{gloss:Base Model}{base model} and \glosslink{gloss:Fine-Tuning}{fine-tune} it once per each task domain.
This would, for example, result in one model that specializes in sentiment classification, another which specializes in automatic summarization, another in part-of-speech tagging, and so on.
Many state-of-the-art systems today handle a wide variety of tasks using a single model.\footnote{It is rumored that the models underlying ChatGPT are actually an \emph{ensemble} of on the order of $10$ expert models, in which different types of requests get routed to specific experts. Nevertheless, if true, these experts are still more flexible than task-specific models from the past.}

These models are able to do all sorts of things~\citep{ganguli2022surprise}. 
In Section~\ref{sec:vision}, we discuss how Generative AI is a generative technology, in the sense of Jonathan Zittrain's theory of generativity~\citep{zittrainfuture}. 
The scaling up of Generative AI (Section~\ref{sec:scale}) has facilitated generativity across a wide range of applications and modalities, not just in the text-to-text and text-to-image applications that are most commonly reported on in the news.
As a non-exhaustive list, this scaling has enabled huge breakthroughs in image captioning~\citep{li2023blip2}, music generation~\citep{agostinelli2023musiclm}, 
speech generation~\citep{le2023voicebox} and transcription~\citep{radford2022whisper}, tools for lowering the barrier to learning to program \citep{yilmaz2023coding}, and research questions in the physical sciences (including on protein folding, drug design, and materials science) \citep{{corso2023diffdock}}.
\vspace{-.2cm}
\notocsubsection{Developments in the Training Pipeline: Pre-Training and Fine-Tuning}\label{sec:dividing}
\vspace{-.1cm}

Machine-learning models are \glosslink{gloss:Training}{trained} on a \glosslink{gloss:Datasets}{training dataset} of \glosslink{gloss:Examples}{examples} of the task that the \glosslink{gloss:Model}{model} is supposed to be able to accomplish.
The nature of these training datasets has changed drastically over the years, leading to the capabilities seen in generative-AI systems today~\citep{lee2023explainers}.
In particular, we have seen a shift toward multi-stage training pipelines, in which models are first trained on large (but possibly lower-quality) datasets to create a \glosslink{gloss:Base Model}{base model} and then progressively trained on smaller, more-curated datasets that better align with the model creators' goals.

Typically, a base model is constructed by training on an enormous, often web-scraped dataset, which instills a ``base'' of knowledge about the world within the model.
This step is called \glosslink{gloss:Pre-Training and Fine-Tuning}{pre-training} because it is the training that occurs before the final training of the model.
As described by \citet{CallisonBurch_2023} in his testimony to the U.S. House of Representatives Judiciary Committee, during pre-training, models learn underlying patterns from their input data. 
When pre-training on large-scale data that has a wide variety of information content, 
base models capture abundant, ``general-knowledge'' information. 
For example, \glosslink{gloss:Large Language Model (LLM)}{large language models (LLMs)} learn syntax and semantics, facts (and fictions) about the world, and opinions, which can be used to produce summaries and perform limited reasoning tasks; image-generation models learn to produce different shapes and objects, which can be composed together in coherent scenes.\footnote{Though, notably, not a collage! See Appendix~\ref{app:metaphors} for a discussion of why the metaphor (Section~\ref{sec:metaphors}) of a \glosslink{metaphor:Generations are collages.}{collage} for generative-AI outputs can be misleading.}
Pre-training gives these models the unprecedented flexibility to generate all sorts of outputs through synthesizing information in the input training data.

This flexibility allows \glosslink{gloss:Base Model}{base models} to be re-used in a variety of ways.
For example, to adapt it to more specific tasks and domains, one can further train (i.e., \glosslink{gloss:Pre-Training and Fine-Tuning}{fine-tune}) the base model on domain-specific data (e.g., legal texts and case documents) to specialize the model's behavior (e.g., performing better at legal document summarization).
Alternatively, one might fine-tune the base model to understand a dialog-like format (as ChatGPT has done \citep{chatgpt}).
Pre-training is very expensive, which means it only happens once (or a small handful of times),\footnote{It can cost millions of dollars to train a \glosslink{gloss:Large Language Model (LLM)}{large language model}~\citep{bloom, bloom-training, lee2023talkin}.} but subsequent fine-tuning tends to be much faster (due to the smaller size of the datasets involved), so it can more tractably occur many times.

\custompar{Choosing what is pre-training and what is fine-tuning} Despite our discussion above of what is unique about \glosslink{gloss:Pre-Training and Fine-Tuning}{pre-training and fine-tuning}, it is worth emphasizing that this division is not well-defined. It is predominantly an artifact of choices made regarding training, rather than an essential 
aspect of the training process. 
Both pre-training and fine-tuning are just 
training (though perhaps configured differently). 
The reasons we differentiate between these two stages have to do with how large-scale model training is done in practice; 
the distinction is only meaningful because researchers frequently choose to
divide stages of training along these lines (in turn, ascribing meaning to this division). 
For example, one actor in the \glosslink{gloss:Supply Chain}{supply chain} may release a pre-trained model, a different actor may fine-tune that model and release it as well, and a third actor may fine-tune the already fine-tuned model~\citep{lee2023talkin}.\footnote{Is the third model fine-tuned from a pre-trained model, or a fine-tuned model? This is all just semantics.}
%
Additionally, researchers frame concrete research questions specifically for pre-training or fine-tuning~\citep[e.g.]{longpre2023pretrainers}.

\custompar{Note} During the GenLaw roundtable discussions, it became apparent that legal experts shared some misconceptions about the roles of pre-training and fine-tuning, and that it is therefore important for machine-learning researchers to emphasize the influence of pre-training on generative-AI model capabilities.
Additionally, we re-emphasize our note from the top of  Section~\ref{sec:knowledge} that pre-training is \emph{training}, and not a data preparation stage. 


\vspace{-.1cm}
\notocsubsection{Generative-AI Systems and the Supply Chain}\label{sec:system}

There are numerous decisions and intervention points throughout the system, which extend to elements beyond choices in pre-training and fine-tuning. 
%
Since many actors can be involved in the generative-AI \glosslink{gloss:Supply Chain}{supply chain}, and decisions made in one part of the supply chain can impact other parts of the supply chain, it can be useful to identify each intervention and decision point and think about them in concert.
We defer to \citet[Part I.C]{lee2023talkin} for detailed discussion of the supply chain and the numerous stages, actors, and design choices that it involves. 

These choices affect the quality of the model, both in terms of its characteristics/capabilities and the model's consequent effectiveness. 
For example, consider the intervention point at which the training data is chosen.
Creating a training dataset requires answering questions like: (1) which data examples should be included in the training dataset; (2) where will the data be stored (e.g., on whose servers); (3) for how long will the data be retained; (4) where will the resulting trained model be deployed, etc?\footnote{For more on choices in training data, see Chapter 1: The Devil is in the Training data from \citet{lee2023explainers}} 
Choices made about where the model will be deployed can affect what training data can be used. 
A model training on private user data has a very different privacy-risk profile if such a model were never to leave the user's personal device, compared to if it is be shared across many users' devices.

%



Not all design choices are about models and how they are trained. 
Models are embedded within overarching systems, which consist of many component pieces that both individually and together reflect the outcomes of relevant sociotechnical design decisions~\citep{cooper2021eaamo, cooper2022arpa, gpt4-systemcard, milesblog}. 
There are numerous other intervention points throughout the supply chain, which involve systems-level choices~\citep{lee2023talkin}. 
Such intervention points include prompt input filters, generation output filters, rate limiting (e.g., how many prompts a user can supply in a given time window to a system), access controls, terms of use, use-case policies for APIs,\footnote{Google, Anthropic, OpenAI, and Cohere all have such policies.} user interface (UI) and experience (UX) design (e.g., to guard against over-reliance on generative-AI systems), and so on~\citep{gpt4-systemcard, copilot-safety-filter, milesblog}. Each of these involve design decisions that can have their own legal implications. 

\notocsubsection{The Massive Scale of Generative-AI Models}\label{sec:scale}


Ultimately, the capacity to facilitate ``magical,'' flexible,  open-ended functionality with Generative AI (Section~\ref{sec:openended}) comes from the massive scale at which generative-AI models are trained~\citep{smith2023convnets}. 
State-of-the-art models today are an order of magnitude larger and trained on significantly more data than the biggest models from five years ago. 

Techniques for \glosslink{gloss:Scale}{scaling up} models have demonstrated a uniquely important role in unlocking gen\-er\-a\-tive-AI capabilities. 
This includes research into more efficient \glosslink{gloss:Architecture}{neural architectures} and better machine-learning systems for handling model training and inference at scale~\citep{ratner2019mlsys}.
For example, many experts have studied methods for collecting and curating massive, web-scraped datasets~\citep[e.g.]{lee2023explainers}, as well as the ``emergent behaviors'' of models trained at such such large scales~\citep[e.g.]{wei2023emergent}. 

In spite of these changes, it is worth noting that many techniques used in Generative AI today are not new.
Language models, for example, have existed since at least the 1980s \citep{rosenfeld2000two}.
The difference is that, in recent years, we have figured out how to scale these techniques tremendously
(e.g., modern \glosslink{gloss:Large Language Model (LLM)}{language models} use \glosslink{gloss:Context Window}{context windows} of thousands of input tokens, compared to the 5 to 10 input tokens used by language models in the early 2000s). 
We defer to machine-learning experts to provide more specific details on the methodologies and outcomes of scaling.\footnote{Part II.B 
of \citet{lee2023talkin} has useful discussion and citations on this topic.} 

Finally, one of the implications of scale is that machine-learning practitioners are training fewer state-of-the art models today than were being trained in the past.
When models were small relative to available computing resources, it was common to re-train a machine-learning system several times, changing \glosslink{gloss:Hyperparameter}{hyperparameters} or other configuration details to find the best-quality model.
Today's model scale means the cost of training \textit{just one} state-of-the-art model can be hundreds of thousands or even millions of dollars~\citep{bloom, bloom-training, lee2023talkin}, 
This further incentivizes the push toward general-purpose models described in Section \ref{sec:openended}.




%% file: section/40-taxonomy.tex
\vspace{-.25cm}
\notocsection{A Preliminary Taxonomy of Legal Issues}\label{sec:taxonomy}

One significant outcome of the GenLaw discussions was  progress toward a taxonomy of the legal issues that Generative AI raises.
We say ``progress toward'' because the initial analysis presented here is very much an interim contribution as part of an ongoing project.
The GenLaw workshop was explicitly scoped to privacy and \glosslink{gloss:The Field of Intellectual Property (IP)}{intellectual property (IP)} issues, so this analysis should be considered non-exhaustive, and the omission of other topics is not a judgment that they are unimportant.
Further, we note that not all capabilities, consequences, risks, and harms of Generative AI are \emph{legal} in nature, so this taxonomy is not a complete guide to generative-AI policy.
Other reports have made significant attempts to catalog such concerns~\citep[e.g.]{epic}. 
We instead focus on highlighting the ways in which specifically legal issues may arise.

We begin with an important high-level point: Generative AI inherits essentially \emph{all} of the issues of AI/ML technology more generally. 
This is so because Generative AI can be used to perform a large and increasing number of tasks for which these other types of ML systems have been used.
For example, instead of using a purpose-built sentiment-analysis model, one might simply prompt an \glosslink{gloss:Large Language Model (LLM)}{LLM} with labeled examples of text and ask it to classify text of interest; 
one could use a trained LLM to answer questions with ``yes'' or ``no'' answers (i.e., to perform classification). 
The resulting classifications may or may not be as reliable as ones from a purpose-built model, but insofar as one is using a machine-learning model in both cases, any legal issues raised by the purpose-built model are also present with the LLM.

Further, any crime or tort that involves communication could potentially be conducted using a generative-AI system.
One could use an LLM to write the text used for fraud, blackmail, defamation, or spam, or use an image-generation system to produce deepfakes, obscene content, or false advertisements.
Almost any speech-related legal issue is likely to arise in some fashion in connection with Generative AI.\looseness=-1

With these broader observations in mind, in the remainder of this section we discuss four legal areas that will need to deal with Generative AI: intention in torts and criminal law (Section~\ref{sec:torts}), privacy (Section~\ref{sec:privacy}), misinformation and disinformation (Section~\ref{sec:badinformation}), and intellectual property (Section~\ref{sec:ip}). 

\notocsubsection{Intent}\label{sec:torts}

Numerous aspects of law turn on an actor's intention.
For example, in criminal law,  the defendant's ``criminal intent'' (\emph{mens rea}), not just the act and its resulting \glosslink{gloss:Harm}{harms}, is often an element of a crime \citep{mensrea}. Intent is not a universal requirement. 
Some crimes and torts are ``strict liability'' (e.g., a manufacturer is liable for physical harm caused by a defective product regardless of whether they intended that harm, which they almost always did not).
But where intent is required, a defendant's lack of wrongful intent means they cannot be convicted or held liable.
For example, fraud is an intentional tort and crime.
A defendant who speaks falsely but honestly to the best of their knowledge does not commit fraud.



Generative AI will force us to rethink the role of intent in the law.
In contrast to most prior types of ML systems,\footnote{Highly autonomous vehicles are another such example, but currently remain in relatively limited use~\citep{cooper2022avs}.} Generative AI 
can cause harms that are similar to those brought about by human actors but \emph{without} human intention.
For example, an LLM might emit false and derogatory claims about a third party -- claims that would constitute defamation if they had been made by a human \cite{volokh2023llms}.


There is unlikely to be a simple across-the-board answer as to how the ``intent'' of a generative-AI system should be measured, in part because the legal system uses intention in so many ways and so many places.
Consider an example from a GenLaw discussion.
One participant noted that it may be useful to  move to a \emph{respondeat superior} model --- a legal doctrine (often used in \glosslink{gloss:Tort}{tort} law) that ascribes the legal responsibility of an employee to their employer (if the tort or other wrongful conduct was conducted within the scope of employment).  
For this kind of liability model, one could treat the generative-AI system as the ``employee,'' and then ascribe responsibility for harm to the ``employer'' -- i.e., the user.
Such an approach appears to sidestep the need to deal with intent;  \emph{respondeat superior} is strict liability as to the employer.
However, another participant noted that in the usual application of \emph{respondeat superior}, there is still an embedded notion of intention.
That is because the employee's intentions are still relevant in determining whether a tort has been committed at all; only if there has can liability then also be placed on the employer.
This is not to say that \emph{respondeat superior} has no role to play, only that it does not avoid difficult questions of intent.

Another line of discussion at GenLaw concerned whether these difficulties might lead to a greater focus on the human recipients of generative-AI outputs.
Some authors, for example, have argued that the rise of AI systems creates a world of ``intentionless free speech'' in 
which communications should be assessed purely based on their utility to the listener \cite{collins2018robotica}.
Such a framework helps establish clearly a First Amendment basis for a right \emph{to use} Generative AI.
But it also raises difficult questions about how to protect users \emph{from} Generative AI in cases of false or harmful outputs.
These issues will cut across many legal areas.


\notocsubsection{Privacy}\label{sec:privacy}

As discussed above (Section~\ref{sec:knowledge}), ``privacy'' is a notoriously difficult term to define. 
Different disciplines rely on different definitions that simplify the concept in different ways,  which can make it very difficult to communicate about privacy across fields. 
In particular, computer science and law are known to operate using very different notions of privacy.\footnote{Even subfields vary greatly. 
Cryptographers use different conceptions of ``privacy'' than ML researchers; decisional privacy in constitutional law is very different than data privacy in technology law.}

For example, in many subfields of computer science, it is common to employ definitions of privacy based on mathematical formalisms that are computationally tractable (often \glosslink{gloss:Differential Privacy}{differential privacy}).
In contrast, privacy in the law is often defined contextually, based on social norms and reasonable expectations.
It is typically necessary to first identify which norms are at play in a given context, after which it is then possible to determine if those norms have been violated (and what to do about it). 
Such definitions of privacy are fundamentally nuanced; they resist quantification. 
The tensions between legal and computer-science approaches to privacy are a source of communication challenges. 
At GenLaw, one of the legal experts provided a useful intuition for this tension: Computer scientists often want to be able to quantify policy, including policy for handling privacy concerns; in the law, the mere desire to quantify complex concepts like privacy can itself be the source of significant problems. 

Despite these difficulties, it is still important to be able to reason about privacy (and when it is \glosslink{gloss:Privacy Violation}{violated}) in both computing and law. 
This is not a new problem: it has been a source of significant practical and research challenges for essentially as long as computers have been in use. 
As long as \glosslink{gloss:Personally Identifiable Information (PII)}{personally identifiable information (PII)} like addresses and phone numbers has been stored on computers, there have been risks that such information could be seen or leveraged by others who otherwise would not have had access.
Since the introduction of machine-learning methods to software systems, it has become possible to predict user behavior and personal preferences (sometimes with high fidelity); in turn, having access to such arguably private information has opened up the possibility to develop software that relies on this information to guide or manipulate user behavior. 
We understand how difficult these privacy challenges are only because of decades of research in law and computer science.
Legal scholars have articulated the real-world \glosslink{gloss:Harm}{harms} that people can suffer from through misuse of ``private'' information; computer scientists have demonstrated real-world attack vectors through which ``private'' information can be leaked.

Generative AI is poised to make these privacy challenges even harder. 
As noted above (Section~\ref{sec:scale}), in contrast to prior machine-learning models, generative-AI models are typically trained on large-scale web-scraped datasets. 
These datasets can contain all sorts of private information (e.g., \glosslink{gloss:Personally Identifiable Information (PII)}{PII})~\citep{brown2022privacy}, which in turn can be memorized and then leaked in \glosslink{gloss:Generation}{generations}~\citep{carlini2023extracting, somepalli2022diffusion}.
A traditional search engine only locates individual data points, but a generative-AI model could link together information in novel ways that reveal sensitive information about individuals. 
Adversarially designed \glosslink{gloss:Prompt}{prompts} can extract other sensitive information, such as internal instructions used within chatbots~\citep{edwards2023leak}.\footnote{This is potentially also a trade-secret issue (Section~\ref{sec:ip}), depending on the nature of the information leaked.} 


\notocsubsection{Misinformation and Disinformation}\label{sec:badinformation}

Generative AI can be used to produce plausible-seeming but false content at scale.
As such, it may be a significant source of misinformation, and amplify the speech of actors engaged in disinformation campaigns.\footnote{By misinformation, we mean material that is false or misleading, regardless of the intent behind it.
In contrast, disinformation consists of deliberately false or misleading material, often with the purpose of manipulating human behavior.}
These capabilities will present issues in any area of law that prohibits false speech --- from lies about people to lies about products to lies about elections.
They will also challenge the assumptions of areas of law that tolerate false speech out of the belief that such speech will be comparatively rare and easy to counter.
Taken together, these include a wide variety of legal topics, including defamation, national security, impersonation, bad (or, in regulated contexts, illegal) advice, over-reliance~\citep{milesblog}, amplification, spear-phishing, spam, elections, consumer-protection law (e.g., addiction, deception, false advertising, products liability), deepfakes, and much else. 

From a misinformation perspective, generative-AI models are very sensitive to their \glosslink{gloss:Datasets}{training data}, which may itself include misinformation or disinformation. 
For example, in August of this year, it was discovered that a book about mushroom foraging, which was produced with the assistance of an \glosslink{gloss:Large Language Model (LLM)}{LLM}, contained misinformation about which mushrooms are poisonous (likely due to inaccurate information learned from the data on which the LLM was trained)~\citep{404mushroom}.
Similarly, it appears that LLMs are subject to \emph{sycophancy}, where a model answers subjective questions in a way that flatters their user's stated beliefs, and \emph{sandbagging}, where models are more likely to endorse common misconceptions when their user appears to be less educated'' \cite{bowman2023eight, perez2022discovering}.

From a disinformation perspective, models can be used 
to deliberately \glosslink{gloss:Generation}{generate} persuasive but false content scraped from the Internet.
But they can also be deliberately manipulated through adversarially selected \glosslink{gloss:Pre-Training and Fine-Tuning}{fine-tuning} data or through \glosslink{gloss:Alignment}{alignment}.
These processes could be used to skew models to deliberately produce misleading content. 



One point raised by legal scholars at GenLaw is that generated disinformation about individuals (e.g., deepfakes) will potentially contribute to new types of defamation-related harms. 
Other experts, particularly those with a privacy background in computing, questioned whether such harms could also be classified as intimate privacy violations~\citep{citron2022privacy}.
In many respects, the harms caused by sufficiently convincing forgeries are very similar to those caused by truthful revelations~\citep{zipursky2023tort}.
Indeed, this is something that Generative AI seems well-positioned to enable: large-scale, inexpensive production of 
believable deepfakes that use a person's likeness, and  depict fake intimate acts or convey fake intimate information~\citep{maiberg2023aiporn, deepfakehs}.

In response, lawyers with expertise in defamation said they believed that this would likely not constitute a cognizable \emph{privacy} \glosslink{gloss:Harm}{harm}  under the law, although it would still be actionable as defamation or false light.
In turn, this response raised questions about whether Generative AI could create new types of harms that blur current conceptions of disinformation and privacy harms. 

\notocsubsection{Intellectual Property}\label{sec:ip}

Naturally, given the recent spate of lawsuits about \glosslink{gloss:Copyright}{copyright} and Generative AI (as well as the stated thematic focus for the first GenLaw Workshop), \glosslink{gloss:The Field of Intellectual Property (IP)}{IP} was a frequent topic for emerging legal issues. 
It has also been one of the first generative-AI subjects explored in detail by scholars, \citep[e.g.]{lee2023talkin, sag2023safety, samuelson, CallisonBurch_2023, vyas2023provable}, and we leave discussion of the doctrinal details to their work.
Instead, we focus here on a few high-level observations about current and impending IP issues --- many of which also have applications beyond IP.
\begin{itemize}
    \item \emph{Volition}: Human volition plays an important and subtle role in defining IP infringement.
    For example, copyright infringement normally requires that a human \emph{intentionally made} a copy of a protected work, but not that the human was consciously aware that they were infringing.
    Generative-AI systems may occasionally produce outputs that look like duplicates of the training data.
    Some participants at GenLaw were concerned that it may be easy to deflect the role of human-made design choices (Section~\ref{sec:dividing}) by making such choices seem ``internal'' to the system (when, in fact, such choices are typically not foregone conclusions or strict technical requirements~\citep{cooper2022accountability}). 
    Since  ``purely internal'' copies tend to be fair use, such deflection could serve as a copyright  liability shield. 
    Legal experts will need to contend with this possibility in their analysis of generative-AI systems.\footnote{For a more general treatment of ``scapegoating the system,'' see \citet{cooper2022accountability}. Its opposite, in which a human is held wholly responsible for a harm caused by a technical system, is the ``moral crumple zone,'' described in \citet{elish2019moral}.} 
    Additionally, since generative-AI systems typically take in input from human users, it is also possible for a user to intentionally cause a model to output potentially infringing content.
    \item \emph{Market externalities}: There are many concerns that Generative AI will lead to mass labor displacement, significant market changes, and the concentration of market power. These issues extend beyond IP, but necessarily invoke related questions of ownership.
    These are matters for labor law, international trade law, and other areas of law.
    But they are also IP issues, because doctrines such as fair use invite courts to consider such societal effects in weighing the propriety of particular copying.
    \item \emph{Trade secrecy}: Fine-tuning on proprietary data is poised to become a potentially useful pattern in the adoption of generative-AI technology. However, existing generative-AI models are known to \glosslink{gloss:Memorization}{memorize} their training data~\citep{carlini2023extracting, carlini2023quantifying}. In turn, this raises the possibility that an adversarial user could extract proprietary information in training data, thereby presenting issues related to trade secrecy~\citep{edwards2023leak}. 
    \item \emph{Scraping}: 
    Similarly, the legality of scraping training data is intextricable from the IP treatment of Generative AI.
    Generative-AI companies both rely on scraped data as an input and take measures (both technical and legal) to prevent outputs from their systems from being used as inputs to other systems without permission.
    \item \emph{Authorship}: As alluded to above, IP law may need to reconsider authorship eligibility in light of Generative AI. Computer authorship is not a new topic of analysis in the law~\citep[e.g.]{grimmelmann2016authored, samuelson1985allocating}, but Generative AI is likely to present new variations on old themes. For example,  purely computer-generated works are not currently covered by copyright. However, some argue that this situation not sustainable~\citep[e.g.]{lee2023author}. Where AI-generated works have significant value, there will be strong economic pressures on courts to gives users  copyright in those works.
    \item \emph{Patent}: Given that generative-AI modeling techniques also have applications in the physical sciences (e.g., in drug design, see Section~\ref{sec:challenges}), it seems likely that there will be implications for \glosslink{gloss:Patent}{patent} law. For example, U.S. patent law requires a human inventor as a condition of patent eligibility. Just as copyright's human authorship requirement has been challenged (but so far upheld \citep[e.g.]{thaler}), similar challenges may arise with respect to patents.
    \item \emph{Idea-expression dichotomy}: Generative AI seems to further blur the already often-murky line between idea and expression in copyright law. For example, one could attempt to analogize the \glosslink{gloss:Prompt}{prompt} to an idea and the associated \glosslink{gloss:Generation}{generation} to its expression, but this presents several problems. For one thing, there seems to be a bit of an inversion from the typical pattern: it suggests that the AI, rather than the human, is responsible for the creative expression (which is not currently protectable by copyright law). For another, there may be sufficient creativity for copyrightability of the prompt itself, even if it is ultimately (by the prior analogy) responsible for the idea in the resulting generation. Lastly, there is a tenable argument that the human prompter and generative-AI system are acting in concert to produce the resulting generation~\citep{lee2023talkin}, and that the way that an idea is expressed in a prompt makes it intextricably indivisible from the resulting expressive generation. In short, as others have noted~\citep{lemley2023upsidedown}, Generative AI seems to turn the idea-expression dichotomy ``upside down.'' 
\end{itemize}

%% file: section/50-agenda.tex
\notocsection{Toward a Long-Term Research Agenda}\label{sec:agenda}

Participants in the GenLaw workshop and roundtable identified several important and promising future research directions. 
Notably, these topics have several elements in common.
First, each showcases how technical design choices play a crucial role in legal research questions.
Many of the architectures and applications of generative-AI systems are genuinely novel, compared to previous technologies that the legal system has had to contend with.
Understanding the legal issues that they raise will require close engagement with the technical details.

Second, just as design choices can inspire questions for the legal scholars, it is also important to consider how legal scholarship can influence the choices that generative-AI researchers make when designing systems (Sections~\ref{sec:dividing},~\ref{sec:system}). 
Understanding not just the current legal framework, but also how that framework may evolve, provides important guidance for system designers about which technical changes are and are not legally significant.
In addition, a clear sense of the legal possibility space can help direct generative-AI research toward novel designs, algorithms, attacks, and characterizations that have beneficial characteristics.

The list that follows is just a  sample of emerging research areas at the intersection of Generative AI and law. 
It gives a flavor of how these two disciplines can concretely inform each other. 
We believe it is the starting point of a rich, long-term research agenda with the potential to influence and inform public policy, education, and industrial best practices.


\notocsubsection{Centralization and Decentralization}

One crucial question about the future of Generative AI concerns the relative degree of centralization versus decentralization.
Consider, as an example, the controversies over the use of \glosslink{gloss:Closed Dataset}{closed}-\glosslink{gloss:License}{licensed} data (within web-scraped datasets) as training data for generative-AI models (e.g., LAION datasets~\citep{laion, laionpaper}, The Pile~\citep{pile}, Books3~\citep{books3wired}, etc.), especially if training involves removing copyright management information. While such datasets are often released with \glosslink{gloss:Open Dataset}{open} licenses (e.g., the LAION organization has released their datasets under the MIT license, which allows for use and copying), this does not guarantee that the associated and constituent data examples in those datasets can be licensed for use in this way~\citep{lee2023talkin}. Many examples within datasets have closed licenses.\footnote{As \citet{lee2023talkin} notes, this is particularly complex for datasets used to train \glosslink{gloss:Multimodal}{multimodal} models, like text-to-image models; the examples to train text-to-image models are image-caption \emph{pairs}, where for each pair the image and the text caption could be subject to their own copyrights (and even hypothetically could be subject to a copyright as a compilation).} 
Legal scholars have made arguments that run the gamut of possible fair-use outcomes for the use of these datasets in Generative AI~\citep[e.g.]{samuelson, lemley2023upsidedown, lee2023talkin, sobel2021taxonomy, sag2023safety, henderson2023foundation}. Nevertheless, it remains to be seen whether courts will rule that the use of such datasets constitutes fair use. 

In the interim, an alternative path is to invest in producing open, permissively licensed datasets that avoid the alleged legal issues of using web-scraped data. This means not only releasing datasets with such licenses, but ensuring that the underlying data examples in the dataset have clear provenance~\citep{lee2023explainers} and are openly licensed.
This is a rich problem domain.
It involves significant technical innovation, both in techniques for collecting such datasets at scale while respecting licensing conditions and also in training models that make best use of the limited materials available in them. 
(Current attempts to train models on such openly licensed datasets have yielded mixed results in terms of generation quality~\citep{gokaslan2023commoncanvas}.)
It also requires substantial legal innovation, including the development of appropriate licenses that function as intended across jurisdictions, and organizational innovation in creating authorities to steward such datasets.

These same issues and tensions recur at every stage in the development of generative-AI systems.
This is partly a technical question; current methods require centralized pre-training at scale based on datasets typically gathered from highly decentralized creators.
Whether either or both of these constraints will change in the future is an important and open question.
Improvements in training algorithms may reduce the investment required to pre-train a powerful \glosslink{gloss:Base Model}{base model}, opening it up to greater decentralization.
At the same time, improvements in synthetic data may enable well-resourced actors to generate their own training data, partially centralizing the data-collection step.

Centralization versus decentralization is also partly a business question (Section~\ref{sec:business}).
There is currently substantial investment both in large centralized companies that are developing large base models, and in a large ecosystem of smaller entities developing fine-tuned models or smaller special-purpose models.
The relationships among, and relative balance between, these different entities is likely to evolve rapidly in the coming years.

And, most significantly, centralization versus decentralization is a fundamentally legal question.
As noted above, licensing law may inform who can use a dataset.
Competition and antitrust law are likely to play a major role going forward.
Every important potential bottleneck in Generative AI -- from copyright ownership to datasets to compute to models and beyond -- will be the focus of close scrutiny.
These novel markets will require technical, economic, and legal analysis to determine the most appropriate competition policy.
In addition to antitrust enforcement, possible policies include government subsidies, open-access requirements, ``public option'' generative-AI infrastructure, export restrictions, and structural separation.
These questions cannot be discussed intelligently without contributions from both technical and legal scholars.


\notocsubsection{Rules, Standards, Reasonableness, and Best Practices}

Since the technological capabilities of today's generative-AI systems are so new, it is unclear what duties the creators and users of these systems should be.
This overarching problem is not unprecedented for either law or computing. 
In some cases, these duties take bright-line rule-like forms; HIPAA strictly regulates which kinds of data are treated as personally identifying and subject to stringent security standards.
In other cases, these are more flexible standards that require greater exercises of discretion.
In some cases, the legal system defaults to a general standard of reasonableness: did a person  behave reasonably when developing or using a system?
And sometimes, even when there is no law on point, practitioners have developed best practices that they follow to do their jobs effectively.
We anticipate that the legal system will need to articulate these duties for generative-AI creators and users, and to determine which modalities of rules and standards to employ.

These expectations have always been technology-specific and  necessarily change over time as technology evolves. 
For example, under the Uniform Trade Secrets Act (UTSA) information must be ``the subject of efforts that are reasonable under the circumstances to maintain its secrecy.''
The threshold for what efforts are considered ``reasonable'' has changed over time in response to developments in information security.
Similarly, in cybersecurity, the FTC monitors the state of the art.
As state-of-the-art practices improve, the FTC has been willing to argue that companies engage in unfair and deceptive trade practices by failing to implement widely used cost-effective measures.
Further, the definition of reasonableness is contextual; what is considered reasonable for a large company (e.g., in terms of system development practices) is typically different than what is considered reasonable for smaller actors.

In short, legal scholars urgently need to study -- and technical scholars urgently need to explain -- which generative-AI safety and security measures are recognized as efficient and effective.
Nor will a one-time exchange suffice.
The legal system must be attuned to the dynamism of generative-AI development.
What is currently an effective countermeasure against extracting memorized examples from models may fail completely in the face of a newly developed techniques.
But conversely, new techniques of \glosslink{gloss:Pre-Training and Fine-Tuning}{training} and \glosslink{gloss:Alignment}{alignment} may be developed that are so clearly effective that it is appropriate to expect future generative-AI creators to employ them.
Indeed, the legal system must be attuned to this dynamism itself, to the fact that our current understanding of the frontier between the possible and the impossible in Generative AI is provisional and constantly being refined.
There is work here for many researchers from both communities.


Once technology begins to stabilize, it becomes easier to define concrete standards (e.g., safety standards).  Accordingly, by definition, compliance with such standards is sufficient for meeting the bar of reasonableness.  
Until there is some stability, when harms occur, there will necessarily be some flexibility; there will be some deference to system builder's self-assessments of whether their design choices reflected reasonable best efforts to construct safe systems. In turn, today's best efforts will guide future standards-setting and determination of best practices. 

Both the legal and machine-learning research communities should face this reality head-on; they should take hold of the opportunity to actively engage in research and public policy regarding today's generative-AI systems, such that they can help shape the development of future standards. 
This work will require understanding the complexity and particulars of different generative-AI technologies; effective standards will differ by model \glosslink{gloss:Multimodal}{modality} and other system capabilities (e.g., generative-AI systems that interact with APIs to bring in additional content, such as plugins~\citep{chatgpt-plugins}). 

To meet this challenge, one clear need is useful metrics to effectively evaluate the behaviors of generative-AI systems.
As we discuss below (Section~\ref{sec:eval}), effective ways to evaluate generative-AI systems currently remain elusive. System capabilities and harms are not readily quantifiable; designing useful metrics will be an important, related area of research for Generative AI and law. 

\notocsubsection{Notice and Takedown $\neq$ Machine Unlearning}\label{sec:unlearning}

Notice and takedown is well-known in both software and legal communities because of search engines and Section 512 of the US Copyright Act.
Enabling notice-and-takedown functionality has a variety of sociotechnical challenges, which are notably even more complicated for Generative AI. 

For generative-AI models, there is no straightforward analogue for simply\footnote{Notice and take down can be technically challenging (but nevertheless feasible) for large-scale software systems that involve distributed databases that work in concert.} removing a piece of data from a database (as might be the case for removing a file from a video-hosting platform).
Once a model has been trained, the impact of each data example in the training data is dispersed throughout the model and cannot be easily traced. 
In order to remove an example from a trained model, one must either track down all the places where the example has an impact and identify a way to negate its influence, or re-train the entire model. 
This is challenging because ``impact'' is not well defined, and neither is ``removal.''\footnote{Alternatively, we must first define what it means to ``take down'' a training example from a generative-AI model, which itself is a ill-defined problem.} 

There are entire subfields of machine learning devoted to problems like these. 
For example, the subfield of ``machine unlearning''~\cite{bourtoule2021machine,cao2015towards} attempts to define the desired goals for removing an example and to design algorithms that satisfy these goals.\footnote{This subfield is also heavily motivated by the \glosslink{gloss:The Right to be Forgotten}{The Right to be Forgotten} clause in \glosslink{gloss:The General Data Protection Regulation (GDPR)}{GDPR}.} Another line of work attempts to quantify data-example attribution and influence; it seeks to  define ``attribution'' and then attribute generations from a model to specific data examples in the training data. Both machine unlearning and attribution are very young fields, and their strategies are (for the most part) not yet computationally feasible to implement in practice for deployed generative-AI systems. 
Machine unlearning and attribution are of significant interest for ML researchers and practitioners. There has been intense (and growing) investment in this area. How these fields will develop remains to be seen.


\notocsubsection{Evaluation Metrics}\label{sec:eval}

Evaluation is far from a new topic in machine learning (or computing more generally). 
Nevertheless, there is a clear need for useful metric definitions for Generative AI. 
We discuss some issues of interest below. 

\custompar{Defining metrics for evaluation}
It is well-known that the force of legal rules depends on how they are implemented and interpreted.
Many decisions are made on a case-by-case basis, taking into account specific facts and context. 
In contrast to this approach, machine-learning practitioners evaluate systems at scale. 
It is common practice to define metrics that can be applied directly to every situation (or at least a large majority of them). 
These metrics necessarily use a pre-specified sets of features that may leave out considerations that may be important to forming a decision that appropriately accounts for broader context. 

This is hardly a new observation; it has had significant influence machine-learning subfields, such as algorithmic fairness. More generally, the challenges of operationalizing or concretizing societal concepts into math has been discussed at length in prior works~\citep[e.g.]{jacobs2021measurement, friedman1996bias, cooper2021emergent, cooper2021eaamo}, and developing reasonable definitions for legal concepts is an active and evolving area of research~\citep[e.g.]{cooper2022lawless, cooper2023variance, scheffler2022formalizing}. 

Nevertheless, it is worth emphasizing that these observations hold true for Generative AI.\footnote{There are also specific complexities for Generative AI that have not been so readily apparent in prior work in machine learning (e.g., in other areas of machine learning, there are accepted (though imperfect) notions of ``ground truth'' labels, which are absent in Generative AI).} 
For example, as we discussed in Section \ref{sec:metaphors}, researchers create different, precise, definitions of memorization for different purposes. 
The 
definition of memorization for an image-generation model will differ greatly from a code-generation model or a text-generation model. 
Similarly, since ``removing'' the impact of a training example from a trained model is an ill-defined problem, researchers may develop different metrics for quantifying whether or not training data points are successfully removed.

\custompar{Evaluation is dynamic}
Many metrics are defined in terms of technical capabilities. For example, the evaluation of the amount of memorized training data in a model depends on the ability to extract and discover the memorized training data~\citep{carlini2023quantifying, ippolito2023preventing}. As the techniques for data extraction improve, the evaluation of the model will change. 
Additionally, the way models are used alters the way that they should be evaluated.
For example, a machine-unlearning method may be applied to a model to remove the effect of a specific individual's data. 
However, that model may later be fine-tuned on additional data that is very similar to the removed individual's data. This may cause the individual's data to effectively ``resurface.''\footnote{This is presently speculation, though  not all-together baseless. Recent research shows that the effects of \glosslink{gloss:Alignment}{alignment} methods may be negated through the course of \glosslink{gloss:Pre-Training and Fine-Tuning}{fine-tuning}~\citep{qi2023finetuning}.}
The way that the \glosslink{gloss:Supply Chain}{supply chain} is constructed for a particular generative-AI model may alter the way that the systems (in which that model is embedded) can and should be evaluated. 
For example, the analysis may change depending on whether a model is \glosslink{gloss:Alignment}{aligned} or not.\footnote{Further, alignment is not binary. There are different possible degrees of alignment.}
In turn, as another example, it is possible that some actors may not have the relevant information to perform necessary evaluations;
some actors may not even know if a particular model is aligned or not. 

%% file: section/60-conclusion.tex
\vspace{-.2cm}
\section{Conclusion and the Future of GenLaw}\label{conclusion}
\vspace{-.2cm}

In this report, we discussed the main topics broached at the first GenLaw workshop: the importance of developing a shared knowledge base for improved communication (Section \ref{sec:knowledge}), the unique aspects of Generative AI that present novel challenges and opportunities (Section~\ref{sec:challenges}), a taxonomy of emerging legal issues (Section~\ref{sec:taxonomy}), and associated open research questions at the intersection of Generative AI and law (Section~\ref{sec:agenda}). 


As is clear from the diversity of issues discussed within these topics, it is difficult to pithily sum up the main takeaways of GenLaw. 
Nevertheless, we will attempt to do so, and will 
sketch out our hopes for the future of GenLaw as an organization. 

\begin{enumerate}
    \item \emph{Expanding beyond copyright concerns}:  Perhaps the fairest overarching assessment from this report is that GenLaw's participants believe that copyright concerns just scratch the surface of potential issues. Put differently, a common belief was that the legal questions currently under consideration in U.S. courts only touch on a small area of the potential legal issues that Generative AI will raise. In part, this is because the underlying technology is continuing the evolve and be adopted at such a rapid pace. As a result, the research agenda that we suggest here (Section~\ref{sec:agenda}) will necessarily evolve over time. 
    \item \emph{Shaking off disciplinary boundaries}:  There remain major open questions about how best to evaluate the behavior of generative-AI systems. Answering these questions necessarily will involve machine-learning-technical knowledge, but they will also involve much more. This report illustrates just how  central legal considerations are to effective evaluation. But we do not intend to suggest that accounting for these considerations will on its own be sufficient. As we continue to understand how Generative AI will transform our interactions, expectations, economy, education, etc., we will need to continue to shake off disciplinary boundaries in order to design useful and comprehensive evaluation methodologies. 
    \item \emph{Evolving resources and engagement}: Given the generative and evolving nature of generative-AI systems and products, GenLaw's work to help educate and facilitate engagement between technologists, legal experts, policymakers, and the general public will necessarily require ongoing effort. The  resources that we develop (such as those in this report) will need to be frequently updated to keep pace with technological changes. 
\end{enumerate}

%

%

In response to these takeaways, we are growing GenLaw into a nonprofit\footnote{GenLaw is in the process of obtaining 501(c)(3) status.} home for research, education, and interdisciplinary discussion.
Thus far, we have written pieces that make complex and specialized knowledge about law and Generative AI accessible to both a general audience~\citep{lee2023explainers} and to subject-matter  experts~\citep{lee2023talkin}. We have worked to provide additional resources, such as recordings of our events~\citep{livestream, liveblog}, collections of external resources~\citep{resources}, and the initial glossary on the GenLaw website. 
%
For our first in-person workshop, we engaged with participants that have various expertise in Generative AI, law, policy, and other computer-science disciplines across 25 different institutions.
We are excited to continue engaging with experts across industry, academia, and government. 
While our first event and materials have had a U.S.-based orientation, we are actively focusing on expanding our engagement globally. 
We will be maintaining the GenLaw website\footnote{\url{https://genlaw.github.io}} with the most up-to-date information about future events and resources. 

%% file: section/99-appendix.tex
\input{section/appendix/10-glossary}



%% file: section/appendix/10-glossary.tex
\notocsection{Glossary}\label{app:glossary}

We roughly divide our glossary into two sections: terms from machine learning and Generative AI (Appendix \ref{gloss:machine-learning} - \ref{app:os}), and those from the law (Appendix \ref{gloss:legal-concepts} - \ref{app:privacy}). 
We further subdivide these sections, and then alphabetize terms within them.\\

\newlist{glossarylist}{enumerate}{1}
\setlist[glossarylist]{
label={},
topsep=0pt,
leftmargin=2em,
listparindent=0em,
itemindent=-2em,
parsep=0.5em,}

\newcommand{\glossitem}[1]{\item \hypertarget{gloss:#1}{\nummeriere*[#1]}\textbf{\textcolor{glosscolor}{#1}}\hspace{1em}}
\newcommand{\metaphoritem}[1]{\hypertarget{metaphor:#1}{\paragraph*{\textcolor{glosscolor}{#1}}}}

\startcontents[sections]
\contentsuse{paragraph}{psections}
\setcounter{secnumdepth}{5}
\setcounter{tocdepth}{5}
\printcontents[sections]{l}{3}{}

\newpage

\subsection{Concepts in Machine Learning and Generative AI}\label{gloss:machine-learning}
\begin{glossarylist}

\glossitem{Algorithm}
An \textbf{algorithm} is a formal, step-by-step specification of a process. 
%
Machine learning uses algorithms for \glosslink{gloss:Training}{training} \glosslink{gloss:Model}{models} and for
applying models (a process called \glosslink{gloss:Inference}{inference}). 
In training, the algorithm takes a model \glosslink{gloss:Architecture}{architecture}, \glosslink{gloss:Datasets}{training data}, \glosslink{gloss:Hyperparameter}{hyperparameters}, and a
random seed (to enable random choices during statistical computations) to produce trained model \glosslink{gloss:Parameters}{parameters}. 

In public discourse around social media, the term algorithm is often used to refer to methods for optimizing the probability that a user will engage with a post; 
however, it is important to note that algorithms describe many processes including, for example, the process of sorting social media posts by date.

\glossitem{Alignment} \textbf{Alignment} refers to the process of taking a \glosslink{gloss:Pre-Training and Fine-Tuning}{pre-trained} \glosslink{gloss:Model}{model} and further tuning it so that its outputs are \textit{aligned} with a policy set forth by the model developer. 
Alignment can also refer to the \textit{state of being aligned} --- some academic papers might compare and contrast between an aligned and an unaligned model.
The goals included in an alignment policy (sometimes called a constitution) vary from developer to developer, but common ones include:
\begin{itemize}
    \item Following the intent of user-provided instructions; 
    \item Abiding by human values (e.g., not emitting swear-words); 
    \item Being polite, factual, or helpful;
    \item Avoiding generating copyrighted text.
\end{itemize}

While the goals of alignment are vast and pluralistic, current techniques for achieving better alignment are broadly applicable across goals.
These techniques include \glosslink{gloss:Reinforcement Learning}{reinforcement learning} with human feedback and full model \glosslink{gloss:Pre-Training and Fine-Tuning}{fine-tuning}.
Specifying the desired properties of alignment often requires a special dataset. These datasets may include user-provided feedback, supplied through the  user interface in a generative-AI product.

\glossitem{Application Programming Interface (API)}
Companies choose between releasing generative-AI \glosslink{gloss:Model}{model} functionality in a variety of ways: 
They can release the model directly by
\glosslink{gloss:Open Model}{open-sourcing} it; 
they can embed the model in a software system, which they release as a product; 
or, they can make the model (or the system its embedded in) available via an \textbf{Application Programming Interface (API)}. 
When a model is open-source, anyone can take the \glosslink{gloss:Checkpoint}{checkpoint} and load it onto a personal computer in
order to use it for \glosslink{gloss:Generation}{generation} (by embedding the checkpoint in a program). 
In contrast, when a company only makes their generative-AI model or system available via an API, that means that users access it in code. 
The user writes a query in a format specified by the company, then sends the query to the company's server. 
The company then runs the model or system on their own computers, and provides a response to the user with the generated content. API access usually requires accepting the company's \glosslink{gloss:Terms of Service}{Terms of Service}, and
companies may add extra layers of security on top of the model (such as rejecting queries identified as being in violation of the ToS).

\glossitem{Architecture} The design of a \glosslink{gloss:Model}{model} is called its \textbf{architecture}. 
For a \glosslink{gloss:Neural Network}{neural network,} architectural decisions include the format of inputs that can be accepted (e.g., images with a certain number of pixels), the number of layers, how many \glosslink{gloss:Parameters}{parameters} per layer,
how the parameters in each layer are connected to each other, and how we represent the intermediate state of the model as each layer transforms input to output.
The most common architecture for language tasks is called a Transformer~\citep{vaswani2017attention}, for which there are many variations.

Many contemporary models appear in \textbf{model families} that have similar
architectures but different sizes, often differentiated by the total number of parameters in the model. 
For example, Meta originally released four sizes of the
LLaMA family that had almost the exact same architectures, differing only in the number of layers and size of the
intermediate \glosslink{gloss:Vector Representation}{vector representations}.
More layers and wider
internal representations can improve the capability of a model, but can
also increase the amount of time it takes to \glosslink{gloss:Training}{train} the model or
to do \glosslink{gloss:Inference}{inference.}

\glossitem{Attention} An \textbf{attention} mechanism is a sub-component of a \glosslink{gloss:Transformer}{transformer} architecture. 
This mechanism allows a \glosslink{gloss:Neural Network}{neural network} to selectively focus on (i.e., \textbf{attend} to) specific tokens in the input sequence by assigning different attention weights to each token. 
Attention also refers to the influence an input token has on the token that is generated by the model. 

\glossitem{Base Model}
A \textbf{base model}, which is sometimes called a ``\glosslink{gloss:Foundation Model}{foundation}'' or ``pre-trained model,'' is a \glosslink{gloss:Neural Network}{neural network} that has been \glosslink{gloss:Pre-Training and Fine-Tuning}{pre-trained} on a large, general-purpose dataset (e.g., a \glosslink{gloss:Web Crawl}{web crawl}).
Base models can be thought of as the first step of training and a good building block for other models.
Thus, base models are not typically exposed to Generative AI users; instead they are adapted to be more usable either through \glosslink{gloss:Alignment}{alignment} or \glosslink{gloss:Pre-Training and Fine-Tuning}{fine-tuning} or performing \glosslink{gloss:In-Context Learning (Zero-Shot / Few-Shot)}{in-context learning} for \textit{specific} tasks.
For example, OpenAI trained a base model called GPT-3 then adapted it to follow natural-language instructions from users to create a subsequent model called InstructGPT~\citep{instructgpt}.

See also: \glosslink{gloss:Pre-Training and Fine-Tuning}{pre-training and fine-tuning}. 

\glossitem{Checkpoint}
While a \glosslink{gloss:Model}{model} is being trained, all of its \glosslink{gloss:Parameters}{parameters}
are stored in the computer's memory, which gets reset if the program terminates or the computer turns off. 
To keep a model around long-term,
it is written to long-term memory, i.e., a hard drive in a file called a \textbf{checkpoint.} 
Often, during training, checkpoints are written to disk every several
thousand steps of training. The minimum bar for a model to be considered
\glosslink{gloss:Open Model}{open-source} is if there has been a public release of one of its
checkpoints, as well as the code needed to load the checkpoint back into
memory.

\glossitem{Context Window} 
Also called \textbf{prompt length}.
Generative AI \glosslink{gloss:Prompt}{prompts} typically have a fixed \textbf{context window}. This is the maximum accepted input length for the model and arises because models are trained with data examples that are no longer than this maximum context window. 
Inputs longer than this maximum context window may result in generations with performance degradations. 

\glossitem{Data Curation and Pre-Processing}
\textbf{Data curation} is the process of creating and curating a dataset for training a model. 
In the past, when datasets were smaller, 
they could be manually curated, with human annotators assessing the quality of each example. Today, the final datasets used to train generative machine-learning models are typically  automatically curated. 
For example, data examples
identified as ``toxic'' or ``low quality'' may be removed. 
This filtering is done using an \glosslink{gloss:Algorithm}{algorithm}. These algorithms may include heuristic rules (e.g., labeling examples containing forbidden words as toxic) or may use a different machine-learning model trained to classify training examples as low-quality or toxic. 
Another common curation step is to deduplicate the dataset by identifying examples that
are very similar to each other and removing all but one copy~\citep{lee2022dedup}.

\textbf{Data pre-processing} involves transforming each example in the dataset to
a format that is useful for the task we want our machine-learning model
to be able to do. For example, a web page downloaded from a \glosslink{gloss:Web Crawl}{web
crawl} might contain HTML markup and site navigation headers. We may
want to remove these components and keep only the content text.

Data curation and pre-processing happens directly to the data and are
independent of any model. 
Different models can be trained on the same dataset. 
For more information about
dataset curation and pre-processing, see \citet[Chapter 1]{lee2023explainers}


\glossitem{Datasets}
\textbf{Datasets} are collections of data \glosslink{gloss:Examples}{examples}. Datasets are used
to \glosslink{gloss:Training}{train} machine-learning models. This means that the
\glosslink{gloss:Parameters}{parameters} of a machine-learning model depend on the dataset.
Different machine-learning models may require different types of data
and thus different datasets. The choice of dataset depends on the task
we want the machine-learning model to be able to accomplish. For example,
to train a model that can generate images from natural-language descriptions, we would need a dataset consisting of aligned image-text pairs.

Datasets are often copied and reused for multiple projects because (a)
they are expensive and time-consuming to create, and (b) reuse makes it
easier to compare new models and algorithms to existing methods (a
process called \textbf{benchmarking}). Datasets are usually divided into
\glosslink{gloss:Training}{training} and \textbf{testing} portions. The testing portion is
not used during training, and is instead used to measure how well the
resulting model \glosslink{gloss:Generalization}{generalizes} to new data (how well the model performs on examples that look similar to the training data but were not actually
used for training). 

Datasets can be either created directly from a raw source, such as Wikipedia or a \glosslink{gloss:Web Crawl}{web crawl}, or they can be created by assembling
together pre-existing datasets. Here are some popular ``source''
datasets used to train generative machine learning models:

\begin{itemize}
\tightlist
\item
  \href{https://huggingface.co/datasets/wikipedia}{Wikipedia}
\item
  \href{https://arxiv.org/abs/1812.08092v1}{Project Gutenberg}
\item
  \href{https://commoncrawl.org/}{Common Crawl}
\item
  \href{https://github.com/allenai/allennlp/discussions/5056}{C4}
\item
  \href{https://www.image-net.org/}{ImageNet}
\item
  \href{https://laion.ai/blog/laion-400-open-dataset/}{LAION-400M}
\item
  \href{https://huggingface.co/papers/2303.03915}{ROOTS}
\end{itemize}

Here are some popular datasets that were created by collecting together
several pre-existing datasets:

\begin{itemize}
\tightlist
\item
  The Pile (The Pile was formerly listed online, but was removed following the Huckabee v. Meta Platforms, Inc., Bloomberg L.P., Bloomberg Finance, L.P., Microsoft Corporation, and The EleutherAI Institute class action complaint \cite{huckabee}. Notably, a de-duplicated version of the dataset is still available on HuggingFace via EleutherAI, as of the writing of this report~\citep{deduppile}.)
\item
  \href{https://blog.allenai.org/dolma-3-trillion-tokens-open-llm-corpus-9a0ff4b8da64}{Dolma}
\item
  \href{https://github.com/togethercomputer/RedPajama-Data}{RedPajama}
\end{itemize}

Collection-based datasets tend to have a separate license for each
constituent dataset rather than a single overarching license.

\glossitem{Decoding}
A \textbf{decoding algorithm} is used by a \glosslink{gloss:Language Model}{language model} to generate the next word given the previous words in a prompt. There are many different types of decoding algorithms including: greedy algorithm, beam-search, and top-k~\citep{decoding}.
For language models, the term \textbf{decoding} may be used interchangeably with \glosslink{gloss:Inference}{inference}, \glosslink{gloss:Generation}{generation}, and \textbf{sampling}. 

\glossitem{Diffusion-Based Modeling} 
\textbf{Diffusion-based modeling} is an algorithmic process for model training. 
Diffusion is \emph{not} itself a \glosslink{gloss:Architecture}{model architecture}, but describes a process for training a model architecture (typically, an underlying \glosslink{gloss:Neural Network}{neural network})~\citep{sohldickstein2015dpm, rombach2022diffusion, ho2020denoising, song2019grads}. 
Diffusion-based models are commonly used for image generation, such as in Stable Diffusion text-to-image models~\citep{stable-diffusion}. \textbf{Diffusion probabilistic model}, \textbf{diffusion model}, and \textbf{latent diffusion model} have become terms that refer to types of models (typically, neural networks) that are trained using diffusion processes. 


\glossitem{Embedding} 
(Also called \textbf{vector representation}.)
\textbf{Embeddings} are numerical representations of data. 
There are different types of embeddings, such as word embeddings, sentence embeddings, image embeddings, etc.~\citep[p. 26, pp. 703-10]{murphy2022mlintro}.
Embeddings created with machine-learning models seek to model statistical relationships between the data seen during training. 

``Common embedding strategies capture semantic similarity, where vectors with similar numerical representations (as measured by a chosen distance metric) reflect words with similar meanings.

Needless to say, such quantified data are not identical to the entities they reflect, \emph{however}, they can capture certain useful information about said entities'' Quoted from~\citet[p. 7]{lee2023talkin}:

\glossitem{Examples}
An \textbf{example} is a self-contained piece of data.
Examples are assembled into \glosslink{gloss:Datasets}{datasets}.
Depending on the dataset, an example can be an image, a piece of text (such as content of a web page), a sound snippet, a video, or some combination of these.
Often times examples are \textbf{labeled} --- they have an input component that can be passed into a machine-learning model, and they have a target output, which is what we would like the model to predict when it sees the input.
This format is usually referred to as an ``input-output'' or ``input-target'' pair.
For example, an input-target example pair for training an image-classification model would consist of an image as the input, and a label (e.g.~whether this animal is a dog or a cat) as the target.

\glossitem{Generalization} 
\textbf{Generalization} in machine learning refers to a model's ability to perform well on unseen data, i.e., data it was not exposed to during training. 
\textbf{Generalization error} is usually measured evaluating the model on \textbf{training} data and comparing it with the evaluation of the model on \textbf{test} data~\citep{generalization}. (See \glosslink{gloss:Datasets}{datasets} for more on training and test data.)

\glossitem{Generation}
\textbf{Generative models} produce complex, human-interpretable outputs,
such as full sentences or natural-looking images, called
\textbf{generations}. 
Generation also refers to the process of applying the generative-AI model to an input and generating an output. 
The input to a generative model is often called a \glosslink{gloss:Prompt}{prompt}.
More traditional machine-learning models are limited to ranges of numeric outputs (\textbf{regression}) or discrete
output labels like ``cat'' and ``dog'' (\textbf{classification}).
More commonly the word \glosslink{gloss:Inference}{inference} is used to describe applying a traditional machine-learning model to inputs.

Generative models can output many different generations for the same prompt, which all may be valid to a user.
For example, there may be many different kinds of cats that
would all look great wearing a hat. This makes evaluating the performance of
generative models challenging (i.e., how can we tell which is the ``best'' such cat?). 
Recent developments in Generative AI have significantly increased generation quality. 
However, 
even high-quality generations can reach an
\textbf{uncanny valley} with subtly wrong details (e.g., incorrect number of fingers on a typical human hand). 

See also: \glosslink{gloss:Inference}{inference}.

\glossitem{Fine-Tuning}
See description alongside \glosslink{gloss:Pre-Training and Fine-Tuning}{pre-training}.

\glossitem{Foundation Model}
\textbf{Foundation model} is a term coined by researchers at Stanford University \cite{bommasani2021foundation} to refer to neural networks that are trained on very large general-purpose datasets, after which they can be adapted to many different applications.
Another common word, used interchangeably with foundation model, is \glosslink{gloss:Base Model}{base model}.

\glossitem{Hallucination}
There are two definitions of the word \textbf{hallucination}~\citep{hallucination}.
First, hallucination can refer to generation that does not accord with our understanding of
reality, e.g.,  
the generation for the text prompt ``bibliographies'' consisting
of research papers that do not exist. These hallucinations may occur
because generative models do not have explicit representations of facts
or knowledge. Second, generations that have nothing to do with the input
may also be termed a hallucination, e.g., an image generated of a
cat wearing a hat in response to the prompt ``Science fiction from the
80s.''

\glossitem{Hyperparameter}
\textbf{Hyperparameters} are settings for the model or training process that are manually specified  prior to training. 
Unlike \glosslink{gloss:Parameters}{parameters}, they are not typically learned (often, they are hand-selected). 
Examples of hyperparameters for a model include properties of the
architecture, such as input sequence length, the number of model parameters per \textbf{layer}, and number of layers. 
Examples of hyperparameters that determine the behavior of the training algorithm include the learning rate, which controls how much we update model parameters after each input/output training example (i.e., the magnitude of the update).
The process of picking hyperparameters is typically called \textbf{hyperparameter optimization}, which is its own field of research. 

\glossitem{In-Context Learning (Zero-Shot / Few-Shot)}
A \glosslink{gloss:Base Model}{base model} can be used directly without
creating new \glosslink{gloss:Checkpoint}{checkpoints} through \glosslink{gloss:Pre-Training and Fine-Tuning}{fine-tuning}. 
\textbf{In-context learning} is a method of adapting the model to a specific application by providing additional \glosslink{gloss:Context Window}{context} to the model through the \glosslink{gloss:Prompt}{prompt}. 
This can be used instead of the more computationally expensive \glosslink{gloss:Pre-Training and Fine-Tuning}{fine-tuning} process, though it may not be as effective.
In-context learning involves creating an input or \glosslink{gloss:Prompt}{prompt} to a model in a way that
constrains a desired output. Typically the input includes either
\textbf{instructions} (\textbf{zero-shot}: a natural language description of the output) or
a small number of \textbf{examples} of input-output pairs (\textbf{few-shot}). 
``\textbf{Shot}'' refers to the number of examples provided.

\glossitem{Inference} More traditional machine-learning methods (like \textbf{regression} and \textbf{classification}) often use the word \textbf{inference} instead of \glosslink{gloss:Generation}{generation} to refer to the process of applying a trained model to input data.
However, both \textbf{inference} and \textbf{generation} are used to describe the generation process for a generative-AI model.

See also: \glosslink{gloss:Generation}{generation}.

\glossitem{Language Model}
\hypertarget{language-model}{}
A \textbf{language model (LM)} is a type of \glosslink{gloss:Model}{model} that takes a sequence
of text as input and returns a prediction for what the next word in the
text sequence should be. This prediction is usually in the form of a
probability distribution. For example, when passed the input ``It is
raining,'' the language model might output that the probability of
``outside'' is 70\%, and the probability of ``inside'' is 5\%. Language
models used to be entirely statistical; the probability of a ``outside'' coming next in the phrase would be computed by counting the number of times ``outside'' occurred after
the sequence ``It is raining'' in their training \glosslink{gloss:Datasets}{dataset}.
Modern language models are implemented using \glosslink{gloss:Neural Network}{neural networks}, which have
the key advantage that they can base their output probabilities on
complex relationships between sequences in the training data, rather
than just counting how often each possible sequence occurs.
As an illustrative example, a neural network may be able to use information from a phrase like ``It's a monsoon outside'' occurring in the training data to increase the probability of the word ``outside.''

A language model can be used for \glosslink{gloss:Generation}{generation} by employing an
\glosslink{gloss:Algorithm}{algorithm} that selects a word to generate given the
probabilities output by the model for some prompt sequence. After
each word is selected, the algorithm appends that word to the previous
\glosslink{gloss:Prompt}{prompt} to create a new prompt, which is then used to pick the next word,
and so on. Such an algorithm is referred to as a \glosslink{gloss:Decoding}{decoding
algorithm}.

Language-model generation can be used to implement many tasks, including
autocomplete (given the start of a sentence, how should the sentence be
completed?) and translation (translate a sentence from English to
Chinese). The probabilities output by the language model can also be
used directly for tasks. For example, for a sentiment classification
task, we might ask the language model whether ``but'' or ``because'' is a more
probable next word given the prompt ``the food was absolutely
delicious.''

For more information, see \citet{riedl2023transformers}. 

\glossitem{Large Language Model (LLM)}
The term \textbf{large language model (LLM)} has become popular as a way to distinguish older \glosslink{gloss:Language Model}{language models} from more modern ones,  which use the \glosslink{gloss:Transformer}{transformer} architecture with many parameters and are trained on web-scale datasets.
Older models used different \glosslink{gloss:Architecture}{model architectures} and fewer \glosslink{gloss:Parameters}{parameters}, and were often trained on smaller, more narrowly scoped datasets.
Consensus for what constitutes ``large'' has shifted over time as previous generations of large language models are replaced with models with even more parameters that are trained for even more steps.

\glossitem{Loss}
\glosslink{gloss:Neural Network}{Neural networks} (and other models) take an input and predict an output.
The distance (measured by some specified function) between this output and the output we \textit{expect} the model to predict (i.e., the target) is called the \textbf{loss}.
Neural networks are \glosslink{gloss:Training}{trained} by an \glosslink{gloss:Algorithm}{algorithm} that repeatedly passes \glosslink{gloss:Examples}{examples} into the the network, measures the loss compared to the expected output, and then updates the network's \glosslink{gloss:Parameters}{parameters} so as to reduce the size of the loss on these examples.
The goal of this training process is to minimizing the loss over all the exampling in the training dataset.

Some research areas (for example, \glosslink{gloss:Reinforcement Learning}{reinforcement learning}) refer to ``maximizing a reward'' rather than ``minimizing a loss.''
These concepts are largely interchangeable; a loss can be turned into a reward by adding a negative sign to it.

See also: \glosslink{gloss:Objective}{objective}.

\glossitem{Memorization} 
\textbf{Memorization} generally refers to being able to deduce or produce a \glosslink{gloss:Model}{model's} given training \glosslink{gloss:Examples}{example}.

There are further delineations in the literature about different types of memorization.
A training \glosslink{gloss:Examples}{example} may be \textbf{memorized} by a model if information
about that training example can be \textbf{discovered} inside the model
through any means. 
A training example is said to be \textbf{extracted}
from a model if that model can be prompted to generate an
output that looks exactly or almost exactly the same as the training
example. A training example may be \textbf{regurgitated} by the model if
the generation looks very similar or almost exactly the same as the
training example (with or without the user's intention to extract that
training example from the model). 

To tease these words apart: a
training example is \textbf{memorized} by a model and can be
\textbf{regurgitated} in the generation process regardless of whether the
intent is to \textbf{extract} the example or not.

The word memorization itself may be used to refer to other concepts that
we may colloquially understand as ``memorization.'' For example, facts
and style (artists style) may also be memorized, regurgitated, and
extracted.
However, this use should not be confused with technical words (e.g., \textbf{extraction}) with precise definitions that correspond to metrics.  

\glossitem{Model}
\textbf{Models} are at the core of contemporary machine learning. A model is a mathematical tool that takes an \textbf{input} and produces an \textbf{output}. A simple example might be a model that tells you whether a temperature is above or below average for a specific geographic location. In this case the input is a number (temperature) and the output is binary (above/below). 

There could be many versions of this model depending on geographic location. The behavior of the model is defined by an internal \glosslink{gloss:Parameters}{parameter} (or, \glosslink{gloss:Weights}{weight}). In our temperature example, the model has one parameter, the average temperature for the location. The process for setting the value of the parameter for a specific version of the model is called \glosslink{gloss:Training}{training}. In this case we might train a model for New York City by gathering historical temperature data and calculating the average. The process of gathering historical data is called \glosslink{gloss:Data Curation and Pre-Processing}{data collection}. The process of training --- in this case, calculating the average --- is an \glosslink{gloss:Algorithm}{algorithm}. 

A saved copy of a model's trained parameters is called a \glosslink{gloss:Checkpoint}{checkpoint}. We might save separate checkpoints for different cities or save new checkpoints for our New York City model if we retrain with new data. The process of applying the model to new inputs is called \glosslink{gloss:Inference}{inference}. To create an output for a temperature, we also apply an algorithm: subtract the parameter from the input, and return ``above'' if the difference is positive. 

Our temperature example is a very simple model. In machine learning, models can be arbitrarily complex. A common type of model \glosslink{gloss:Architecture}{architecture} is a \glosslink{gloss:Neural Network}{neural network}, which (today) can have billions of parameters. 

It is important to note that models are often embedded within \textbf{software systems} that can get deployed to public-facing users. For example, GPT-4 is a generative-AI model that is embedded within the ChatGPT system, which also has a user interface, developer \glosslink{gloss:Application Programming Interface (API)}{APIs}, and other functionality, like \textbf{input and output filters}. The other components are not part of the model itself, but can work in concert with the model to provide overall functionality. 

\glossitem{Multimodal}
\hypertarget{multimodal}{}
Generative-AI models may generate content in one \textbf{modality} (text, images,
audio, etc.) or in \textbf{multiple modalities}. For example, DALL-E is a
multimodal model that transforms text to images.

\glossitem{Neural Network}
\hypertarget{neural-network}{}
A \textbf{neural network} is a type of \glosslink{gloss:Architecture}{model architecture}. 
%
%
Neural networks consist of \textbf{layers}, where the output of one layer is used as the input of the next layer. 
Each layer consists of a set of classifiers (\textbf{neurons}) that each performs a simple operation independently of one another.
A neural-network model synthesizes multiple simple decisions by passing the input through a series of intermediate transformations.
The outputs of all classifiers at layer
$n$ are then passed to each classifier in layer $n+1$, and so
forth. Each classifier in each layer has \glosslink{gloss:Parameters}{parameters} that define
how it responds to input. 

\glossitem{Objective}
\textbf{Objective} is a term of art in machine learning that describes a mathematical function for the \textbf{reward} or \glosslink{gloss:Loss}{loss} of a model. 
A common point of confusion is between the \textbf{goal} a model creator might have and the \textbf{objective} for a model.
The goal of model training is typically to create models that \glosslink{gloss:Generalization}{generalize} well. 
But that goal (creating models that generalize well) is not a mathematical function that can be maximized.  As an example, the training objective for \glosslink{gloss:Transformer}{transformer} models is typically to generate the same next token as in the training data. 


\glossitem{Parameters}
\textbf{Parameters} are numbers that define the specific behavior of a \glosslink{gloss:Model}{model}.  For example, in the linear equation model $y=mx+b$, there are two parameters: the slope $m$  and the \textbf{bias} (or,  \textbf{intercept}) $b$. A more complex example might be a model that predicts the probability a person makes a bicycle trip given the current temperature and rainfall. This could have two parameters: one representing the effect of temperature and one representing the effect of rainfall. 
Contemporary \glosslink{gloss:Neural Network}{neural network} models have millions to billions of parameters. Model parameters are often interchangeably referred to as model \glosslink{gloss:Weights}{weights}. The values of parameters are saved in files called \glosslink{gloss:Checkpoint}{checkpoints}.

For more on the distinction between \textbf{parameters} and \textbf{hyperparameters}, see \glosslink{gloss:Hyperparameter}{hyperparameters}.

\glossitem{Pre-Processing}
See \glosslink{gloss:Data Curation and Pre-Processing}{data curation and pre-processing}.

\glossitem{Pre-Training and Fine-Tuning}
Current protocols divide \glosslink{gloss:Training}{training} into a common \textbf{pre-training} phase that results in a general-purpose or \glosslink{gloss:Base Model}{base model} (sometimes called a \glosslink{gloss:Foundation Model}{foundation} or pre-trained model) and an application-specific \textbf{fine-tuning} phase that adapts a pre-trained model \glosslink{gloss:Checkpoint}{checkpoint} to perform a desired task using additional data. 
This paradigm has become common over the last five years, especially as \glosslink{gloss:Architecture}{model architectures} have become larger and larger.
This is because, relative to pre-training datasets, fine-tuning datasets are smaller, which make fine-tuning faster and less
expensive. 
In general, it is much cheaper to fine-tune an existing base model for a particular task than it is to train a new model from scratch.

As a concrete example, fine-tuning models to ``follow instructions'' has become an important special case with the popularity of ChatGPT (this is called \textbf{instruction tuning}). Examples of
interactions in which someone makes a request and someone else follows
those instructions are relatively rare on the Internet compared to, for
example, question-and-answer forums. As a result, such datasets are often
constructed specifically for the purpose of \glosslink{gloss:Large Language Model (LLM)}{LLM} fine-tuning,
and may provide substantial practical benefits for commercial companies. 

Because pre-trained models are most useful if they provide a good basis
for many distinct applications, model builders have a strong incentive
to collect as much pre-training data from as many distinct sources as
possible. Fine-tuning results in a completely new model checkpoint
(potentially gigabytes of data that must be loaded and served separately
from the original model), and tends to require hundreds to thousands of
application-specific examples.

However, the distinction between pre-training and fine-tuning is not well-defined. 
Models are often trained with many (more than two) training stages. 
For example, the choice to call the first two of, say three, training stages pre-training and the last stage fine-tuning is simply a choice. 
Finally, pre-training should not be confused with \glosslink{gloss:Data Curation and Pre-Processing}{data curation or pre-processing}.

\glossitem{Prompt}
Most generative-AI systems take as input (currently, this is often some text), which is then used to condition the output.
This input is called the \textbf{prompt}.

\glossitem{Regurgitation} See \glosslink{gloss:Memorization}{memorization}.

\glossitem{Reinforcement Learning}
\hypertarget{reinforcement-learning}{}
\textbf{Reinforcement learning (RL)} is a type of machine learning for incorporating feedback into systems. 
For generative models, RL is commonly used to incorporate \textbf{human feedback (HF)} about whether the \glosslink{gloss:Generation}{generations} were ``good'' or ``useful,'' which can then be used to improve future generations.
For example, ChatGPT collects ``thumbs-up'' and ``thumbs-down'' feedback on interactions in the system.
User feedback is just one way of collecting human feedback. 
Model creators can also pay testers to rate generations. 
 
\glossitem{Reward} See \glosslink{gloss:Loss}{loss}.



\glossitem{Scale}\textbf{}
Machine-learning experts use the term \textbf{scale} to refer to the number of \glosslink{gloss:Parameters}{parameters} in their model, the size of their training datasets (commonly measured in terms of number of \glosslink{gloss:Examples}{examples} or storage size on disk), or the computational requirements to train the model.
The scale of the model's parameter count and the training dataset size directly influence the computational requirements of training.
The scale of computation needed for training can be measured in terms of the number of GPU-hours (on a given GPU type), the number of computers/GPUs involved in total, or the number of FLOPs (floating point operations).

Machine-learning practitioners will sometimes talk about \textbf{scaling up} a model.
This usually means figuring out a way to increase one of the properties listed above.
It can also mean figuring out how to increase the fidelity of training examples, e.g., training on longer text sequences or on higher-resolution images. 

\glossitem{Supply Chain}
\hypertarget{supply-chain}{}
Generative-AI systems are created, deployed, and used in a variety of different ways. Other work has written about how it is useful to think of these systems in terms of a \textbf{supply chain}
that involves many different stages and actors. For more information
about the generative-AI supply chain, please see \citet{lee2023talkin}.

\glossitem{Tokenization} 
For \glosslink{gloss:Language Model}{language models}, a common \glosslink{gloss:Data Curation and Pre-Processing}{pre-processing} step is to break documents into segments called \textbf{tokens}. For example, the input ``I like ice cream.''
might be tokenized into {[}``I'', ``like'', ``ice'', ``cream'',
``.''{]}. 
The tokens can then be mapped to entries in a
\textbf{vocabulary}.
Each entry, or token, in the vocabulary is given an ID (a number representing that token). 
Each token in the vocabulary has a corresponding \glosslink{gloss:Embedding}{embedding} that is a learned \glosslink{gloss:Parameters}{parameter}.
The embedding turns words into numeric representations that can be interpreted and modified by a model.

Each model family tends to share a vocabulary, which is optimized to represent a particular training corpus. 
Most current models use \textbf{subword tokenization} to handle words that would otherwise not
be recognized. Therefore, a rare or misspelled word might be represented
by multiple tokens, for example {[}``seren'', ``dipity''{]} for
``serendipity.'' The number of tokens used to represent an input is
important because it determines how large the effective \glosslink{gloss:Context Window}{context window} of a
model is.

Tokens are also used for other modalities, like music. Music tokens may be \textbf{semantic tokens} that may be created using another \glosslink{gloss:Neural Network}{neural network}.

\glossitem{Transformer}
A \textbf{transformer} is a popular \glosslink{gloss:Architecture}{model architecture} for
image, text, and music applications, and the transformer architecture
underlies models like ChatGPT, Bard, and MusicLM. An input (text or
image) is broken into segments (word \glosslink{gloss:Tokenization}{tokens} or image patches)
as a \glosslink{gloss:Data Curation and Pre-Processing}{pre-processing} step. These input segments are then passed through a
series of layers that generate \glosslink{gloss:Vector Representation}{vector representations} of the
segments. The model has trainable parameters that determine how much
\glosslink{gloss:Attention}{attention} is paid to parts of the input. Like many other
generative-AI models, the transformer model is trained with an \glosslink{gloss:Objective}{objective} that rewards reproducing a target training \glosslink{gloss:Examples}{example}. 

\glossitem{Training}
\hypertarget{training}{}
Machine-learning \glosslink{gloss:Model}{models} all contain \glosslink{gloss:Parameters}{parameters}.
For \glosslink{gloss:Neural Network}{neural networks}, these parameters are typically initialized to random numbers when the network is first created. 
During an \glosslink{gloss:Algorithm}{algorithmic} process called \textbf{training}, these parameters are repeatedly updated based on the training data within the \glosslink{gloss:Datasets}{training dataset} that the model has seen.
Each update is designed to increase the chance that when a model is provided some input, it outputs a value close to the target value we would like it to output. 
By presenting the model with
all of the \glosslink{gloss:Examples}{examples} in a dataset and updating the parameters
after each presentation, the model can become quite good at doing the
task we want it to do.

A common algorithm for training neural network models is
\textbf{stochastic gradient descent,} or SGD. Training datasets are
often too large to process all at once, so SGD operates on small
\textbf{batches} of dozens to hundreds of examples at a time.
Upon seeing a training example, the algorithm generates the model's
output based on the current setting of the parameters and compares that
output to the desired output from the training data. In the case of a \glosslink{gloss:Language Model}{language model}, we might ask: did we correctly choose the next word? 
If the output did not match, the algorithm works backwards through the model's layers, modifying the model's parameters so that the correct output becomes more likely. 
This process can be thought of as leaving ``echoes'' of the training examples encoded in the parameters of the model.

\glossitem{Vector Representation}
See \glosslink{gloss:Embedding}{embedding}.

\glossitem{Web Crawl} A \textbf{web crawl} is a catalog of the web
pages accessible on the Internet. It is created by a \textbf{web crawler}, an
\glosslink{gloss:Algorithm}{algorithm} that systematically browses the Internet, trying to
reach every single web page (or a specified subset). For example, Google Search functions using a
web crawl of the Internet that can be efficiently queried and ranked.
Web crawls are very expensive and compute-intensive to create, and so
most companies keep their crawls private. However, there is one public
web crawl, called \href{https://commoncrawl.org/}{Common Crawl}. Most
open-source (and many closed-source) \glosslink{gloss:Language Model}{language models} are trained at least in part on data extracted from the Common Crawl.

\glossitem{Weights} See \glosslink{gloss:Parameters}{parameters}.

\end{glossarylist}

\subsection{Open versus Closed Software}\label{app:os}

In general, an informational resource --- such as software or data --- is \textbf{open} when it is publicly available for free reuse by others and \textbf{closed} when it is not.
Openness has both practical and legal dimensions.
The practical dimension is that the information must actually be available to the public.
For example, software is open in this sense when its source code is available to the public.
The legal dimension is that the public must have the legal right to reuse the information.
This legal dimension is typically ensured with an ``open-source'' or ``public'' license that provides any member of the public with the right to use the information.

Open versus closed is not a binary distinction.
For one thing, an informational artifact could be practically open but legally closed.
For another, there are numerous different licenses, which provide users with different rights.
Instead, it is always important to break down the specific ways in which information is open for reuse and the ways in which it is not.
The relevant forms of openness are different for different types of information.
In this section, we discuss some of the common variations on open and closed datasets, models, and software.\\

\begin{glossarylist}
\glossitem{Closed Dataset}
\hypertarget{closed-2}{}
Many \glosslink{gloss:Model}{models}, including \glosslink{gloss:Open Model}{open models}, have been trained on non-public (i.e., \textbf{closed}) \glosslink{gloss:Datasets}{datasets}. Though a high-level description of the dataset may have been released and some portions of it may indeed be public (e.g.~nearly all \glosslink{gloss:Large Language Model (LLM)}{large language models} are trained on Wikipedia), there is insufficient public information for the dataset to be fully reproduced. For example, there might be very little information available on the \glosslink{gloss:Data Curation and Pre-Processing}{curation and pre-processing} techniques applied, or constituent datasets might be described in general terms such as ``books'' or ``social media conversations'' without any detail about the source of these datasets. GPT-4 and PaLM are both examples of models trained on non-public datasets.

\glossitem{Closed Model}
\hypertarget{closed-1}{}
When a \glosslink{gloss:Model}{model} is described as \textbf{closed}, it might mean one of three different things.
First, a model might have been described in a technical report or paper, but there is no way for members of the public to access or use the model.
This is the most closed a model can be. For example, DeepMind described the Chinchilla model in a blog post and paper, but the model was never made accessible to the public~\citep{hoffmann2022training}.
Second, a model's \glosslink{gloss:Checkpoint}{checkpoint} may not be publicly available, but the general public may be able to access the model in a limited way via an \glosslink{gloss:Application Programming Interface (API)}{API} or a web application (often for a fee and with the requirement they must sign a \glosslink{gloss:Terms of Service}{Terms of Service}). For example, OpenAI's GPT-3.5 and GPT-4 have followed this paradigm.
In this case, it might be possible for users to reconstruct some of the model's characteristics, but just as with \glosslink{gloss:Closed Software}{closed-source software}, they do not have access to the full details.
Third, the model itself may have been publicly released for inspection, but without a \glosslink{gloss:License}{license} that allows others to make free use of it.
Practically, the license restrictions may be unenforceable against individual users, but the lack of an open license effectively prevents others from building major products using the model.


\glossitem{Closed Software}
\hypertarget{closed-software}{}
\textbf{Closed-source software} is any software where the source code has not been made available to the public for inspection, modification, or enhancement. It is worth noting that closed software can contain \glosslink{gloss:Open Software}{open} components. For example, an overall system might be \textbf{semi-closed} if it releases its \glosslink{gloss:Open Model}{model}, but does not disclose its \glosslink{gloss:Closed Dataset}{dataset}. 
Many open-source \glosslink{gloss:License}{licenses}, which were developed before the advent of modern generative-AI systems, do not prevent \glosslink{gloss:Open Software}{open-source software} from being combined with closed components in this way.


\glossitem{Open Dataset}
\hypertarget{open-2}{}
Saying that a \glosslink{gloss:Datasets}{dataset} is \textbf{open} or \textbf{open-source} can mean one of several things.
At the most open end of the spectrum, it can mean that the dataset is broadly available to the public for download and that the code and experimental settings used to create it are entirely open-source. 
(This is the fullest expression of the idea that making a software artifact open requires providing access to the preferred form for studying and modifying it.)
In some cases, a dataset is available for download but the code and exact experimental settings used to create it are not public. 
In both these situations, use of the dataset is normally governed by an open-source \glosslink{gloss:License}{license}. 

For example, one popular set of benchmark datasets for \glosslink{gloss:Large Language Model (LLM)}{large language
models} is called SuperGLUE. The licenses for its constituent datasets
include \href{https://people.ict.usc.edu/~gordon/copa.html}{BSD
2-Clause},
\href{https://github.com/google-research-datasets/boolean-questions}{Creative Commons Share-Alike 3.0}, and the
\href{https://github.com/rudinger/winogender-schemas/tree/master}{MIT License}. 
In more restrictive cases, the dataset is public, but users must agree to contractual \glosslink{gloss:Terms of Service}{Terms of Service} to access it --- and those terms impose restrictions on how the dataset can be used.
Finally, many datasets are public but cost money to access. For example, the \href{https://www.ldc.upenn.edu/}{UPenn Linguistics Data Consortium} has a catalog of hundreds of high-quality datasets, but individuals need to be affiliated with a member institution of the consortium to access them.

\glossitem{Open Model}
The machine-learning community has described a \glosslink{gloss:Model}{model} as \textbf{open-source} when a trained
\glosslink{gloss:Checkpoint}{checkpoint} has been released with a \glosslink{gloss:License}{license} allowing anyone to download and use it, and the software package needed to load the checkpoint and perform \glosslink{gloss:Inference}{inference} with it have also been open-sourced. 
A model can be open-sourced even in cases where details about how the model was developed have not been made public (sometimes, such models are referred to instead as \textbf{semi-closed}). 
For example, the model creators may not have open-sourced the software package for training the model --- indeed, they may have not even publicly documented the training procedure in a technical report. 
Furthermore, a model being open-source does not necessarily mean the training \glosslink{gloss:Datasets}{data} has been made public.

However, various pre-existing open communities (including the Open Source Initiative, which maintains the canonical Open Source Definition) have objected to usage by the machine-learning community, arguing that it does not capture several of the most important qualities of openness as it has been understood by the software community for over two decades. These qualities include the freedom to inspect the software, the freedom to use the software for any purpose, the ability to modify the software, and the freedom to distribute modifications to others.
\looseness=-1

\glossitem{Open Software}
\hypertarget{open-software}{}
\textbf{Open-source software} is software with source code that anyone can
inspect, modify, and enhance, for any purpose. Typically, such software is
licensed under a standardized open-source \glosslink{gloss:License}{license}, such as the
\href{https://opensource.org/license/mit/}{MIT License} or the  \href{https://opensource.org/license/apache-2-0/}{Apache
license}. Machine-learning systems typically consist of several relatively independent pieces of software; there is the software that builds the \glosslink{gloss:Datasets}{training dataset}, the software that \glosslink{gloss:Training}{trains} the model, and the software that does \glosslink{gloss:Inference}{inference} with the \glosslink{gloss:Model}{model}. Each of these can be independently open-sourced. 


\end{glossarylist}

\hypertarget{legal-concepts}{%
\subsection{Legal Concepts in Intellectual Property and Software}\label{gloss:legal-concepts}}

\begin{glossarylist}

\glossitem{Claims}
\hypertarget{claims}{}
\glosslink{gloss:Patent}{Patent} \textbf{claims} are extremely precise statements that define the scope of protection within the patent. 
Patent claims are carefully written to be broad enough to encompass potential variations of the invention and specific enough to distinguish the invention from prior art.

\glossitem{Copyright}
\textbf{Copyright} grants exclusive rights to creators of original works. For a work to be
copyrightable, it must meet certain criteria: (1) it must be original,
and (2) it must possess a sufficient degree of creativity. Copyright
does not protect facts or concepts, but expressions of those ideas fixed
in a tangible medium (e.g., the idea for a movie, if not written down or
recorded in some way, is typically not copyrightable; a screenplay is
typically copyrightable). Copyright laws provide protections for various
forms of creative expression, including, but not limited to, literary
works, artistic works, musical composition, movies, and software~\citep{copyright}.

See \glosslink{gloss:Idea vs. Expression}{idea vs. expression}.

\glossitem{Copyright Infringement}
\hypertarget{copyright-infringement}{}
\textbf{Copyright infringement} occurs when someone uses, reproduces,
distributes, performs, or displays copyrighted materials without
permission from the copyright owner. This act breaches the exclusive rights held by the copyright holder.

\glossitem{Damages}
\hypertarget{damages}{} 
In the context of \glosslink{gloss:The Field of Intellectual Property (IP)}{IP}, \textbf{damages} refers to financial compensation awarded to the owner of IP for harms and losses sustained as a result of IP infringement.
When IP rights such as \glosslink{gloss:Patent}{patents}, \glosslink{gloss:Copyright}{copyright}, or trademarks are violated, the owner of the IP may file a legal claim for damages. These can cover lost profits, reputational damage, or \glosslink{gloss:License}{licensing} fees.

\glossitem{Fair Use}
\hypertarget{fair-use}{}
\textbf{Fair use} is a legal concept that allows limited use of \glosslink{gloss:Copyright}{copyrighted} materials without permission from the copyright owner~\citep{fairuse}. 
Typically, fair use applies to contexts such as teaching, research, and news reporting, and fair use analyses consider the purpose of use, scope, and the amount of material used.

\glossitem{The Field of Intellectual Property (IP)}
\hypertarget{the-field-of-ip}{}
The field of \textbf{Intellectual Property (IP)} refers to a set of laws that grant exclusive rights for creative and inventive works. 
IP laws protect and promote ideas by providing incentives for innovation and protecting owners of inventions (e.g.~written works, music, designs, among others). 
Intellectual property laws include \glosslink{gloss:Copyright}{copyright}, \glosslink{gloss:Patent}{patents}, trademarks and trade dress, and trade secrets.
Intellectual property law is often the first recourse for conflicts around emerging technologies where more specific legislation has not yet crystallized.
Property law is well-developed, widely applicable, and carries significant penalties including fines and forced removal or destruction of work.

\glossitem{Harm}
\hypertarget{harm}{}
Many areas of law only regulate actions that cause some identifiable \textbf{harm} to specific victims.
For example, people who have not suffered individual harms may not have ``standing'' to bring a lawsuit in federal court.
For another, \glosslink{gloss:Damages}{damage} awards and other \textbf{remedies} may be limited to the harms suffered by the plaintiffs, rather than dealing more broadly with the consequences of the defendant's conduct.
It is important to note that what counts as a cognizable harm is a legal question, and does not always correspond to people's intuitive senses of when someone has suffered a harm.
Physical injury is the most obvious and widely accepted type of legal harm; other widely recognized forms of harm include damage to property, economic losses, loss of liberty, restrictions on speech, and some kinds of \glosslink{gloss:Privacy Violation}{privacy violations}.
But other cases have held that fear of future injury is not a present harm.
Thus, having one's personal information included in a data breach may not be a harm by itself --- but out-of-pocket costs and hassle to cancel credit cards are recognized harms.

\glossitem{Idea vs. Expression}
\hypertarget{idea-vs.-expression}{}
This \textbf{idea vs.~expression} dichotomy gets at the distinction between underlying concepts (or ideas) conveyed by a work, and the specific, tangible manner in which those are expressed. 
An idea refers to an abstract concept or notion behind a creative work, and ideas are not subject to \glosslink{gloss:Copyright}{copyright} protection.
However, expressions, as tangible manifestations, are. 
Tangible fixed expressions of ideas include words, music, code, or art. 
It is important to note that within copyright law, rights are granted to expression of ideas, not ideas themselves. 

See \glosslink{gloss:Copyright}{copyright}.

\glossitem{License}
\hypertarget{license}{}
A \textbf{license} gives legal permission or authorization, granted by the rights holder to others. 
License agreements explicitly outline rights that are granted, as well as limitations, restrictions, and other provisions related to its scope, for example, duration. 
Licenses are common practice within the \glosslink{gloss:The Field of Intellectual Property (IP)}{field of IP}, and are commonly used in software, music, and film industries.

\glossitem{Non-Expressive or Non-Consumptive}
\hypertarget{non-expressive-or-non-consumptive}{}
Certain uses of \glosslink{gloss:Copyright}{copyrighted} materials can be \textbf{non-expressive} or \textbf{non-consumptive}. 
In such cases, copyrighted material is used in a way that does not involve expressing or displaying original work to users. 
Some examples include text mining, building a search engine, or various forms of computational analyses.

\glossitem{Patent}
\hypertarget{patent}{}
A \textbf{patent} confers exclusive rights to inventors, granting them the authority to prevent others from making, using, or selling their inventions without permission. 
Patents create incentives for innovation by providing inventors with a time-based protection from the filing date. 
To obtain a patent, inventions must be new, inventive, and industrially applicable. 
Creators apply for patents; their applications must contain \glosslink{gloss:Claims}{claims} that describe what is novel in the work.

\glossitem{Prior Art}
\hypertarget{prior-art}{}
\textbf{Prior art} is evidence of existing knowledge or information that is publicly available before a certain date. 
Prior art is critical in adjudicating the novelty and nonobviousness of a new invention and may include other \glosslink{gloss:Patent}{patents}. 
Patent examiners search for prior art to determine the patentability of the \glosslink{gloss:Claims}{claimed} invention. Further, prior art informs the patent's applicability and scope.

\glossitem{Terms of Service}
\hypertarget{terms-of-service}{}
\textbf{Terms of service (ToS)} refers to a contractual agreement between the \glosslink{gloss:The Field of Intellectual Property (IP)}{IP} owner and users of \glosslink{gloss:License}{licenses} that govern the use and access to the protected content. 
ToS outline rights, restrictions, and obligations involved. 
ToS may specify permitted uses, licensing terms, and how IP may be copied or distributed. 
ToS safeguard IP owners' rights and ensure compliance with legal standards in the use of IP.

\glossitem{Transformative Use}
\hypertarget{transformative-use}{} 
Expression can build on prior expression. 
In some cases, a new piece of \glosslink{gloss:Copyright}{copyrightable} material may borrow or re-purpose material from prior work. 
If this new material creates something inventive, new, and
substantially different from the original work, then it can be
considered \textbf{transformative use} of the original work, as opposed to \glosslink{gloss:Copyright Infringement}{infringing} on the original copyright owner's exclusive
rights. 
The new material may also be copyright eligible. 
Parody is one common type of transformative use.

\end{glossarylist}

\subsection{Privacy}\label{app:privacy}

\begin{glossarylist}

\glossitem{Anonymization}
\hypertarget{anonymization}{}
\textbf{Anonymization} is the process of removing or modifying personal data in a way that it cannot be attributed to an identifiable individual.

\glossitem{The California Consumer Privacy Act (CCPA)}
\hypertarget{the-california-consumer-privacy-act-ccpa}{}
The \textbf{CCPA} is a California state law that provides consumers with the right to know what personal information businesses collect about them, the right to request their personal information be deleted, and the right to opt-out of sales of their personal information. 
The CCPA applies to all businesses that operate in California, as well as those outside of California that may transfer or process the personal information of California residents.

\glossitem{Consent}
\textbf{Consent} is the voluntary and informed agreement given by an individual for the collection, use, or disclosure of their personal information. 
In the context of data, consent often requires clear and specific communication about the purpose and use of collected data.

\glossitem{Differential Privacy} 
\textbf{Differential privacy} (DP) is an approach for modifying algorithms to protect 
the membership of a given \textbf{record} in a dataset~\citep{dp}. 
Informally, these guarantees are conferred by adding small amounts of \textbf{noise} to the individual data \glosslink{gloss:Examples}{examples} in the dataset. 
Let us say that there are two version of a dataset, $D$ and $D'$, where the former contains an example $E$ and the latter does not. 
If we were to run differentially private \glosslink{gloss:Algorithm}{algorithms} to compute statistics on the datasets $D$ and $D'$, we would not be able to tell by those statistics which dataset contains $E$ and which does not. 
As a result, we can no longer use the computed statistics to infer whether or not the original training data contained the example $E$.

Differential privacy is a theoretical framework that encounters some challenges in practice. 
For example, the amount of noise one must add to data may impact the accuracy of statistics computed, or, when used for generative-AI models, may impact performance of the model.
On the other hand, someone using a differentally private approach needs to add enough noise to ensure that the two datasets $D$ and $D'$ cannot be differentiated through computed statistics. 
Finally, differential privacy was originally created for tabular data and encounters challenges adapting to the unstructured data commonly used for generative-AI models. 
For more on the challenges of applying differential privacy to \glosslink{gloss:Large Language Model (LLM)}{large language models}, please see~\citet{brown2022privacy}.

\glossitem{The General Data Protection Regulation (GDPR)}
\textbf{GDPR} is a comprehensive data protection law implemented by the European Union in 2018~\citep{gdpr}. 
The GDPR governs the collection, use, storage, and protection of personal data for EU residents. 
The law sets out specific rights for individuals regarding their personal data, such as the right to access, rectify, and delete their data, as well as the right to know how their data is being processed.
Further, the GDPR imposes obligations on organizations such as businesses that handle personal data to ensure that proper data protection measures are in place and that \glosslink{gloss:Consent}{consent} is obtained for data processing. 
Non-compliance results in fines and penalties.

\glossitem{Personally Identifiable Information (PII)}
\hypertarget{personally-identifiable-information-pii}{}
\textbf{Personally Identifiable Information (PII)} refers to data that can be used to identify an individual. 
PII can include names, addresses, phone numbers, social security numbers, email addresses, financial information, and biometric data. 
PII is sensitive, and organizations that collect PII are required to implement appropriate measures, adhering to relevant data protection laws (such as the \glosslink{gloss:The General Data Protection Regulation (GDPR)}{GDPR}) to safeguard its confidentiality and integrity.

\glossitem{Privacy Policy}
\hypertarget{privacy-policy}{}
A \textbf{privacy policy} consists of documents that outline how organizations collect, use, store, and protect personal information.
Privacy policies are meant to inform individuals about their rights and the organization's data processing practices.

\glossitem{Privacy Violation}
\hypertarget{privacy-violation}{}
A \textbf{privacy violation} involves unauthorized or inappropriate intrusion into an individual's personal information or activities.
Privacy violations may occur in various forms from data breaches, surveillance, identity theft, or sharing personal or sensitive information without \glosslink{gloss:Consent}{consent}. 
These violations may lead to significant harm such as the loss of personal autonomy, reputational damage, or financial loss.

\glossitem{The Right to be Forgotten}
\hypertarget{right-to-be-forgotten}{}
Some countries' legal systems recognize a \textbf{right to be forgotten} that grants individuals the ability to request the removal of their personal information from online platforms or search-engine results.
The idea is that the legitimate public interest in knowing about other people's past conduct can be outweighed when the information about it is out-of-date or misleading.
The European Union's \glosslink{gloss:The General Data Protection Regulation (GDPR)}{GDPR} includes a form of the right to be forgotten.

\glossitem{Tort}
\hypertarget{tort}{}
A \textbf{tort} is a civil wrongdoing that causes \glosslink{gloss:Harm}{harm} or injury to another person or their property. Tort law provides remedies and compensation to individuals who suffer harm as a result of someone else's actions or negligence.


\end{glossarylist}

\newpage

\hypertarget{Metaphors}{%
\notocsection{Metaphors}\label{app:metaphors}}
In this Appendix, we briefly discuss several metaphors for Generative AI that came up in the GenLaw discussions.
It is worth considering why these metaphors are helpful and where they start to break down.


\metaphoritem{Models are trained.}

Machine-learning practitioners will often say they \glosslink{gloss:Training}{train} \glosslink{gloss:Model}{models}.
Training brings to mind teaching a dog to perform tricks by enforcing good behavior with treats.
Each time the dog performs the desired behavior, they get a treat.
As the dog masters one skill it may move onto another.
Model training is similar in the sense that models are optimized to maximize some \glosslink{gloss:Reward}{reward}.\footnote{Maximizing a reward is exactly equivalent to minimizing a \glosslink{gloss:Loss}{loss} (except for the extra minus sign), but due to historical reasons, machine-learning practitioners use the latter phrasing more often.}
This reward is computed based on how similar the model's outputs are to desired outputs from the model.

However, unlike training a dog, model training does not typically have a curriculum;\footnote{Curriculum learning is an entire field of research in machine learning, but it is not currently standard to use a curriculum.} there is no progression of easier to harder skills to learn, and the formula for computing the reward remains the same throughout model training.


\metaphoritem{Models learn like children do.}
\textbf{Learning} is the active verb we use to describe what a \glosslink{gloss:Model}{model} does as it is being \glosslink{gloss:Training}{trained} ---
a model is \textit{trained}, and during this process it \textit{learns}.
Model learning is the most common anthropomorphic metaphor applied to machine-learning models.
The use of the word \textbf{learning} by machine-learning practitioners has naturally led to comparisons between how models learn and how human children do.
Both children and machine-learning models are ``skilled imitators,'' acquiring knowledge of the world by learning to imitate provided exemplars.  
However, human children and Generative AI obviously use very different mechanisms to learn.
Techniques that help generative-AI systems to learn better, such as increasing model size, have no parallels in child development; mechanisms children use to ``extract novel and abstract structures from the environment beyond statistical patterns'' have no machine-learning comparisons \citep{yiu2023imitation}.

\metaphoritem{Generations are collages.} We quote directly from discussion in~\citet[p. 58]{lee2023talkin}, with added links to our glossary. 
\begin{quote}

It also may seem intuitively attractive to consider \glosslink{gloss:Generation}{generations}  to be analogous to collages.
However, while this may seem like a useful metaphor, it can be misleading in several ways. 
For one, an artist may make a collage by taking several works and splicing them together to form another work. 
In this sense, a generation is not a collage: 
a generative-AI system does not take several works and splice them together.
Instead, \ldots generative-AI systems are built with \glosslink{gloss:Model}{models} \glosslink{gloss:Training}{trained} on many \glosslink{gloss:Examples}{data examples}.
Moreover, those data examples are not explicitly referred back to during the generation process. 
Instead, the extent that a generation resembles specific data examples is dependent on the model \glosslink{gloss:Vector Representation}{encoding} in its \glosslink{gloss:Parameters}{parameters} what the specific data examples look like, and then effectively recreating them.
Ultimately, it is nevertheless possible for a generation to look like a collage of several different data examples;
however, it is debatable whether the the process that produced this appearance meets the definition for a collage.
There is no author ``select[ing], coordinat[ing], or arrang[ing]''\footnote{§ 101 (definition of ``compilation'').} training examples to produce the resulting generation.   
\end{quote}

\metaphoritem{Large language models are stochastic parrots.}
\citet{bender2021parrots} describe a \glosslink{gloss:Large Language Model (LLM)}{large language model} as a stochastic parrot, a
``system for haphazardly stitching together sequences of linguistic forms it has observed in its vast training data, according to probabilistic
information about how they combine, but without any reference to
meaning.''
Like parrots mimicking the sounds that they hear around them, LLMs repeat the phrases they are exposed to, but have no conception of the human meaning behind these phrases.\looseness=-1

This analogy is useful because it references the very real problem of machine-learning models simply outputting their most frequent training data.
Critics of the stochastic-parrot analogy say that it undervalues the competencies that state-of-the-art language models have. Some critics take this further and say that these competencies imply models understand meaning in a human-like way~\citep{piantadosi2022meaning}.\footnote{Whether models are human-like, or the outputs are simply ``really good'' is less pertinent for how generations and inputs should be regulated.}
%
For example, proponents of this analogy might argue that Generative AI passing a difficult standardized exam (such as the Bar Exam \citep{katz2023gpt} or the GRE \citep{gpt4}) is more about parroting training data than human-like skill.
%

\metaphoritem{Large language models are noisy search engines.}

A search engine allows users to search for information within a large database using natural language queries. 
Like a search engine, \glosslink{gloss:Large Language Model (LLM)}{large language models} also return information in response to a natural language query.
However, while a search engine queries the entries in its database and returns the most appropriate ones, a language model does not have direct access to its \glosslink{gloss:Datasets}{training data} and can only make predictions based on the information stored in the model \glosslink{gloss:Weights}{weights}.\footnote{The training data is seen during training, but models are used separately from the training data.}
Most often the output will be a mixture of information contained in many database entries. 
Some model outputs may quote directly from relevant entries in the database (in the case of \glosslink{gloss:Memorization}{memorization}), but this is not reflective of the most typical outputs. 

Sometimes \glosslink{gloss:Generation}{generations} from an LLM will convey similar information that one might learn from running a search; however, sometimes it will not because the underlying \glosslink{gloss:Algorithm}{algorithm} is different. 
Thus, while some generations answer the \glosslink{gloss:Prompt}{prompt} in a similar way to a search, we can more generally think of generative-model outputs as a noisy version of what is actually in the database. Currently, such outputs also tend to lack attribution to the original data entries, and sometimes are incorrect. 
